# Interlayer bond polarizability model for stacking-dependent low-frequency Raman scattering in layered materials

Liangbo Liang,[1,2,*] Alexander A. Puretzky,[1] Bobby G. Sumpter,[1,3] and Vincent Meunier[2,†]

[1]*Center for Nanophase Materials Sciences,*
*Oak Ridge National Laboratory, Oak Ridge, Tennessee 37831, USA*
[2]*Department of Physics, Applied Physics, and Astronomy,*
*Rensselaer Polytechnic Institute, Troy, New York 12180, USA*
[3]*Computational Sciences and Engineering Division,*
*Oak Ridge National Laboratory, Oak Ridge, Tennessee 37831, USA*

## Abstract

Two-dimensional (2D) layered materials have been extensively studied owing to their fascinating and technologically relevant properties. Their functionalities can be often tailored by the interlayer stacking pattern. Low-frequency (LF) Raman spectroscopy provides a quick, non-destructive and inexpensive optical technique for stacking characterization, since the intensities of LF interlayer vibrational modes are sensitive to the details of the stacking. A simple and generalized interlayer bond polarizability model is proposed here to explain and predict how the LF Raman intensities depend on complex stacking sequences for any thickness in a broad array of 2D materials, including graphene, $MoS_2$, $MoSe_2$, $NbSe_2$, $Bi_2Se_3$, GaSe, h-BN, etc. Additionally, a general strategy is proposed to unify the stacking nomenclature for these 2D materials. Our model reveals the fundamental mechanism of LF Raman response to the stacking, and provides general rules for stacking identification.

Notice: This manuscript has been authored by UT-Battelle, LLC under Contract No. DE-AC05-00OR22725 with the U.S. Department of Energy. The United States Government retains and the publisher, by accepting the article for publication, acknowledges that the United States Government retains a non-exclusive, paid-up, irrevocable, world-wide license to publish or reproduce the published form of this manuscript, or allow others to do so, for United States Government purposes. The Department of Energy will provide public access to these results of federally sponsored research in accordance with the DOE Public Access Plan (http://energy.gov/downloads/doe-public-access-plan).

## I. INTRODUCTION

Two-dimensional (2D) layered materials have attracted ever-increasing attention due to their diverse properties of great fundamental and practical interest.[1–9] In 2D materials, the atoms within each layer are joined together by covalent bonds, while much weaker interlayer interactions, mostly van der Waals (vdW) forces, hold the layers together. Consequently, different interlayer stacking configurations can exist, and the stacking order is a powerful approach to tailor the functionalities of 2D materials. For example, in twisted or stacked graphene layers, the stacking change can lead to large Moiré superlattices accompanied by unusual behaviors and new phenomena, such as fractional quantum Hall effects, stacking-dependent Van Hove singularities, etc.[10–15] Graphene trilayers with common ABA (Bernal) and ABC (Rhombohedral) stacking patterns exhibit considerably different electronic structures, infrared absorption, band-gap tunability, and quantum Hall effects.[16–21] In $MoS_2$, the most popular type of layered transition metal dichalcogenides (TMDs), its monolayer exhibits intriguing valley-contrasting optical dichroism for valleytroincs, owing to the strong spin-orbit coupling and broken inversion symmetry.[22–24] In contrast, bilayer $MoS_2$ in the natural 2H stacking restores the inversion symmetry, disabling such optical dichroism.[13,24–26] Proper manipulation of stacking order can break the inversion symmetry and retrieve the strong spin/valley polarizations.[13] $MoS_2$ in the 3R stacking is noncentrosymmetric regardless of the thickness and hence valley-contrasting optical dichroism is always allowed.[13,24,26] The manipulation of the stacking between 2H and 3R can have similar effects on piezoelectricity of $MoS_2$.[27,28]

The precise characterization of stacking is essential to facilitate the efforts in optimizing functional properties of 2D materials. Among many characterization techniques, Raman spectroscopy is a fast, non-destructive, and relatively inexpensive tool that is routinely used in both laboratory and industry.[8,29,30] It has been used for quick identification of the layer thickness and stacking. However, most of previous attempts focused on high-frequency (HF) intralayer modes, which involve vibrations from the intralayer chemical bonds.[4,8,29,31,32] The restoring forces are dominated by the strength of these intralayer chemical bonds, and consequently HF intralayer modes are not very sensitive to the interlayer coupling, which means that there are limitations for them as unambiguous thickness and stacking indicators. In stark contrast, low-frequency (LF) interlayer modes correspond to layer-layer vibrations with each layer moving as a whole unit, and hence their frequencies are solely determined by the interlayer restoring forces and typically below 100 cm$^{-1}$ due to the weak nature of interlayer interactions.[7,9,33] They can be categorized into two types: in-plane shear and out-of-plane breathing vibration modes. Due to their greater sensitivity to interlayer coupling, LF Raman modes can directly probe the interfacial coupling, and they have been found as more effective indicators of the layer thickness[34–46] and stacking[24,47–55] for diverse 2D materials. LF Raman spectroscopy is a rapidly developing field of research and has accelerated due to recent development of volume Bragg gratings for ultra-narrow optical filters with bandwidth about 1 cm$^{-1}$, which allows efficient cut-off of the excitation laser light without employing expensive triple monochromators for LF Raman measurements.[56] The lowest detection limit has been recently pushed down to 2 cm$^{-1}$.[57,58]

In general, the frequencies of LF Raman modes can be used as indicators of thickness. The sensitivity of the frequencies with thickness is now well understood in terms of a linear chain model that treats each layer as a rigid ball and the interlayer coupling as a harmonic spring.[9,30,34,39] Conversely, the intensities of LF Raman modes are sensitive to the stacking and this property can be used to devise stacking indicators. However, a simple model, similar to the linear chain model, does not yet exist for the understanding of the LF Raman intensity dependence on the stacking. Based on an empirical bond polarizability model recently used by Luo et al,[51] we here develop a simplified and generalized interlayer bond polarizability model, which relies on the interlayer bond vectors while omitting atomic coordinates within each layer. The model is mathematically derived in analogy to the linear chain model treating each layer as a single object without structural details. Our model can be applied to both LF shear and breathing modes in diverse stacking sequences for any thickness from bilayer to bulk. We show that the combination of the proposed interlayer bond polarizability model with the linear chain model provides an easy and reliable tool for understanding the thickness and stacking dependence of LF Raman scattering, without requirement of any complicated calculations. Furthermore, in light of inconsistent stacking terminologies used for 2D materials, we propose a general strategy to unify the stacking nomenclature in the field. Overall, our work further facilitates the use of LF Raman spectroscopy for practical identification of both thickness and stacking in 2D materials.

## II. STRUCTURES AND METHODS

### A. Stacking nomenclature for 2D materials

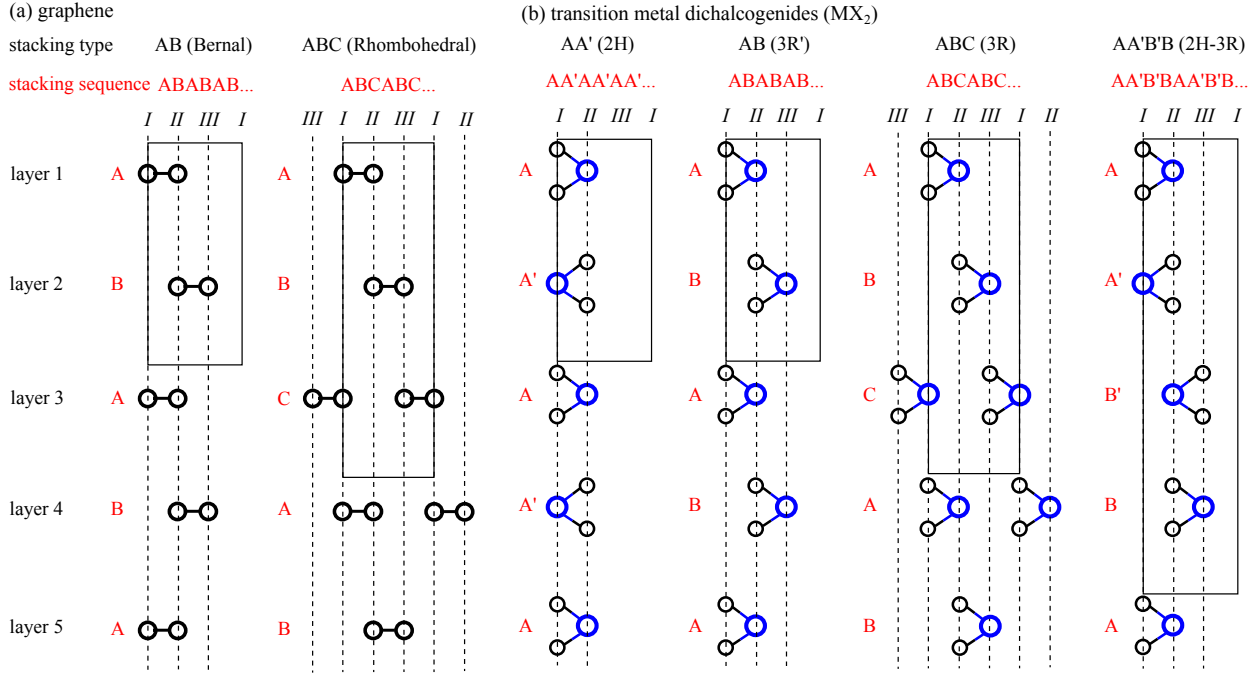

FIG. 1. (Color online) (a) Two major stacking types in graphene and (b) four major stacking types in TMDs $MX_2$ (such as $MoS_2$). In (b), the large blue and small black circles represent the metal M and the chalcogenide X atoms, respectively. In the honeycomb lattice, there are three atomic coordinates on the basal plane: *I*: 0, 0; *II*: 1/3, 2/3; *III*: 2/3, 1/3. Each letter in the stacking sequences stands for one layer. The rectangle with solid lines highlights the unit cell in each stacking.

The most common stacking in bilayer graphene is the AB or Bernal configuration (Fig. 1a), which corresponds to half of the carbon atoms in one layer eclipsed over an atom in the other layer and the other half of the atoms over the hexagon center in the other layer. Compared to the less stable AA stacking where the layers are exactly aligned, AB stacking has the second layer shifted with respect to the first layer, often referred as the *staggered* stacking. In trilayer graphene, the first and second layers assume AB stacking, and for the third layer the shift can be undone to yield an ABA (Bernal) stacking sequence, or repeated to correspond to an ABC (Rhombohedral) stacking sequence (Fig. 1a).[20] AB and ABC are two common stacking types in multilayer graphene, where the former stacking sequence repeats after two layers while the latter repeats after three layers

(Fig. 1a).

For group-6 MX$_2$ like MoS$_2$, the bilayer has two stable stacking patterns, which resemble the bulk 2H and 3R stacking polytypes, respectively.[59–62] Thicker MX$_2$ can assume more complex stacking combinations beyond 2H and 3R.[49,62] Currently different works adopted different terminologies for few-layer MX$_2$, which might create unnecessary confusion in the community.[24,49–51,61–64] This is why the bulk 2H and 3R are still widely used for labeling stacking orders in few-layer systems, though strictly speaking they are not entirely valid and sufficient. For example, for bulk crystals in 2H and 3R configurations, the letters stand for hexagonal and rhombohedral respectively, and the digit indicates the number of layers in one unit cell.[6] Then 3R is not a strictly correct term for bilayer MoS$_2$, as 3R should mean three layers. Of course, the use of 2H and 3R terminology is still recommended as it serves a common ground to start. In this work, we adopted a well-accepted methodology to unify the stacking nomenclature in group-6 TMDs,[49,61–63,65,66] that can be extended to other hexagonal structures including graphene, h-BN, GaS(Se), NbSe$_2$, Bi$_2$Se$_3$, etc.

For bilayer MX$_2$ in Fig. 1b, 2H stacking type corresponds to M in one layer over X in the other layer and X in one layer over M in the other layer (*eclipsed* with M over X), which is named AA$'$ in analogy to the AA stacking in bilayer graphene (*eclipsed* with C over C). 3R stacking corresponds to M in one layer over X in the other layer while all other M and X are over the hexagon centers (*staggered* with M over X), which is assigned to AB because it resembles the AB *staggered* stacking in bilayer graphene (Fig. 1a).[49,61–63,65,66] Note that strictly speaking, 2H stacking cannot be named as AA since AA is reserved for the stacking where M in one layer over M in the other layer and X in one layer over X in the other layer (*eclipsed* with M over M and X over X), the same as for AA stacking in bilayer graphene. AA stacking is unstable and does not exist in natural or synthetic bilayer MX$_2$. According to prior theoretical studies,[63–67] two additional metastable stacking configurations are possible: AB$'$ (*staggered* with M over M) and A$'$B (*staggered* with X over X). In summary, there are five high-symmetry stacking patterns in bilayer MX$_2$ (see Fig. S1 in Supporting Information): two *eclipsed* (AA$'$ and AA) and three *staggered* ones (AB, AB$'$ and A$'$B), among which AA$'$ (2H) and AB (3R) are stable and commonly found in natural and synthetic samples.

Based on AA$'$ and AB stackings, four types of stacking combinations are present in trilayer MX$_2$: AA$'$A, ABA, ABC and AA$'$B$'$ (Fig. 1b). If the first and second layers assume AA$'$ stacking, the second and third layers can assume either A$'$A (equivalent to AA$'$) or A$'$B$'$ (equivalent to AB) stacking, leading to either AA′A or AA′B′ stacking sequence. Obviously, AA′A corresponds to 2H stacking in the trilayer system, while AA′B′ is a mixture of 2H and 3R stacking polytypes (also labeled as 2H-3R). If the first and second layers assume AB stacking, the second and third layers can assume either BB′ (equivalent to AA′) or BA (equivalent to AB) or BC (also equivalent to AB) stacking, leading to either ABB′ or ABA or ABC stacking sequence. ABB′ is also a mixture of 3R and 2H stacking polytypes (labeled as 3R-2H), and it is equivalent to the aforementioned AA′B′ (2H-3R) though inverted. ABA and ABC stacking sequences in trilayer $MX_2$ resemble ABA (Bernal) and ABC (Rhombohedral) stacking sequences in trilayer graphene (Fig. 1). ABC corresponds to the bulk 3R stacking, while ABA is another form of 3R stacking (denoted as 3R′). In short, four non-equivalent stacking types exist in trilayer $MX_2$, and they become AA′AA′, ABAB, ABCA, and AA′B′B in four-layer $MX_2$ respectively (Fig. 1b). AA′ (2H) and AB (3R′) stacking sequences repeat after two layers, ABC (3R) stacking sequence repeats after three layers, while AA′B′B (2H-3R) repeats after four layers. They have been reported in recent works.[24,49–53] The adopted stacking nomenclature here can be extended to group-6 $MX_2$ of any thickness, and it works equally well for h-BN,[68] GaSe[69,70] and $NbSe_2$.[46] All four stacking types can be found in bulk GaSe crystals, where AA′, AB, ABC and AA′B′B are historically named as $\beta$-2H, $\varepsilon$-2H, $\gamma$-3R, $\delta$-4H polytypes, respectively.[69] Additionally, although bulk $NbSe_2$ also assumes the 2H phase, the stacking pattern assumes AB′ (*staggered* with Nb over Nb, while Se atoms over the hexagon centers, Fig. S1),[46] which is different from AA′ stacking of 2H-phase bulk $MoS_2$. This further illustrates the usefulness of the proposed stacking terminology for unambiguous stacking assignment.

### B. Generalized interlayer bond polarizability model

According to the Placzek approximation, the Raman intensity of a phonon mode $k$ is given by[51,71–73]

$$I(k) \propto \frac{n_k+1}{\omega_k} \left| e_i \cdot \tilde{R}(k) \cdot e_s^T \right|^2 = \frac{n_k+1}{\omega_k} \left| \sum_{\mu\nu} e_{i,\mu} e_{s,\nu} \Delta\alpha_{\mu\nu}(k) \right|^2, \quad (1)$$

where $\tilde{R}(k)$ is the (3×3) Raman tensor of the phonon mode $k$, subscripts $\mu$ and $\nu$ indicate Cartesian components ($x$, $y$ or $z$) of the tensor, and $e_i$ and $e_s$ are the unit vectors for the polarization of the incident and scattered light, respectively. $\omega_k$ is the frequency of the phonon mode $k$, and $n_k = (e^{\hbar\omega_k/k_BT} - 1)^{-1}$ is the phonon occupation according to Bose-Einstein statistics. The Raman

tensor elements are

$$\Delta \alpha_{\mu\nu}(k) = \sum_{j\gamma} \left[ \frac{\partial \alpha_{\mu\nu}}{\partial r_{j\gamma}} \right]_0 \Delta r_{j\gamma}(k), \qquad (2)$$

where $r_{j\gamma}$ is the position of atom $j$ along direction $\gamma$ ($x$, $y$ or $z$) in equilibrium, $\left[ \frac{\partial \alpha_{\mu\nu}}{\partial r_{j\gamma}} \right]_0$ is the derivative of the electronic polarizability tensor element $\alpha_{\mu\nu}$ with respect to the atomic displacement from the equilibrium configuration, and $\Delta r_{j\gamma}(k)$ is the eigen-displacement of atom $j$ along direction $\gamma$ in the phonon mode $k$ (i.e., the eigenvector of the mass-normalized dynamic matrix).[72]

One can see that the Raman tensor of the phonon mode $k$ corresponds to the change of the system's polarizability by its vibration. According to the empirical bond polarizability model,[51,73] the polarizability can be approximated by a sum of individual bond polarizabilities from different bonds:

$$\alpha_{\mu\nu} = \frac{1}{2} \sum_{iB} \left[ \frac{\alpha_{\parallel,B} + 2\alpha_{\perp,B}}{3} \delta_{\mu\nu} + (\alpha_{\parallel,B} - \alpha_{\perp,B}) \left( \frac{R_{i\mu,B} R_{i\nu,B}}{R_{i,B}^2} - \frac{1}{3} \delta_{\mu\nu} \right) \right], \qquad (3)$$

where $B$ indicates a bond connected to atom $i$, the boldface $\boldsymbol{R}_{i,B}$ is the corresponding bond vector connecting atom $i$ to one of its neighbor atoms $i'$, $R_{i\mu,B}$ is the $\mu$ ($x$, $y$ or $z$) component of $\boldsymbol{R}_{i,B}$, and $R_{i,B}$ is the length of $\boldsymbol{R}_{i,B}$. $\alpha_{\parallel,B}$ and $\alpha_{\perp,B}$ are the bond polarizabilities for the bond $B$ in the directions parallel and perpendicular to the bond, respectively. After some derivations (details in Supporting Information), the Raman tensor elements are obtained as

$$\begin{aligned}
\Delta \alpha_{\mu\nu}(k) = &-\sum_{iB} \left\{ \hat{\boldsymbol{R}}_{i,B} \cdot \Delta \vec{r}_i(k) \left[ \frac{\alpha'_{\parallel,B} + 2\alpha'_{\perp,B}}{3} \delta_{\mu\nu} + \left( \alpha'_{\parallel,B} - \alpha'_{\perp,B} \right) \left( \hat{R}_{i\mu,B} \hat{R}_{i\nu,B} - \frac{1}{3} \delta_{\mu\nu} \right) \right] \right\} \\
&- \sum_{iB} \left\{ \frac{\alpha_{\parallel,B} - \alpha_{\perp,B}}{R_{i,B}} \left[ \hat{R}_{i\mu,B} \Delta r_{i\nu}(k) + \hat{R}_{i\nu,B} \Delta r_{i\mu}(k) - 2 \hat{R}_{i\mu,B} \hat{R}_{i\nu,B} \left( \hat{\boldsymbol{R}}_{i,B} \cdot \Delta \vec{r}_i(k) \right) \right] \right\},
\end{aligned} \qquad (4)$$

where $\hat{\boldsymbol{R}}_{i,B} = \boldsymbol{R}_{i,B}/R_{i,B}$ is the equilibrium-configuration bond vector normalized to unity, $\hat{R}_{i\mu,B}$ is the $\mu$ ($x$, $y$ or $z$) component of the normalized bond vector, and $R_{i,B}$ is the bond length in equilibrium. $\alpha'_{\parallel,B}$ and $\alpha'_{\perp,B}$ are the radial derivatives of the bond polarizabilities with respect to the bond length.

For an interlayer shear mode vibrating along the $x$ direction, only the $x$ component of $\Delta \vec{r}_i(k)$ can be non-zero, which yields

$$\Delta \alpha_{\mu\nu} = -\sum_{iB} \left\{ \hat{R}_{ix,B} \left[ \frac{\alpha'_{\parallel,B} + 2\alpha'_{\perp,B}}{3} \delta_{\mu\nu} + \left(\alpha'_{\parallel,B} - \alpha'_{\perp,B}\right) \left(\hat{R}_{i\mu,B}\hat{R}_{i\nu,B} - \frac{1}{3}\delta_{\mu\nu}\right) \right] \right.$$
$$\left. + \frac{\alpha_{\parallel,B} - \alpha_{\perp,B}}{R_{i,B}} \left[\hat{R}_{i\mu,B}\delta_{\nu x} + \hat{R}_{i\nu,B}\delta_{\mu x} - 2\hat{R}_{i\mu,B}\hat{R}_{i\nu,B}\hat{R}_{ix,B}\right] \right\} \Delta r_{ix}. \quad (5)$$

Note that for an interlayer vibrational mode in 2D materials, each layer vibrates as an almost rigid body and thus it can be simplified as a single object, where the structural details of each layer can be generally omitted. The bonds within each layer (intralayer bonds) are not compressed/stretched during the interlayer vibration, and thus do not contribute to the change of the polarizability. Only the bonds between the layers (interlayer bonds) are altered during such vibrations, leading to the polarizability change.[51] Subsequently, Eq. 5 can be simplified so that $i$ indicates the index of an entire layer instead of any atom within it, and $B$ indicates a bond connecting layer $i$ to a neighboring layer $i'$ in equilibrium. In general, for layer 1, if the derivative of the system's polarizability with respect to its displacement is defined as $\vec{\alpha}'_1$ and its displacement from the equilibrium position is $\Delta \vec{r}_1$, the change of the polarizability by its displacement is $\vec{\alpha}'_1 \cdot \Delta \vec{r}_1$. Similarly, the change of the polarizability by the displacement of layer 2 is $\vec{\alpha}'_2 \cdot \Delta \vec{r}_2$. The total change of the system's polarizability by the interlayer vibration is a sum of the changes of each layer, which is $\Delta \alpha = \sum_i \vec{\alpha}'_i \cdot \Delta \vec{r}_i = \sum_i (\alpha'_{ix}\Delta r_{ix} + \alpha'_{iy}\Delta r_{iy} + \alpha'_{iz}\Delta r_{iz})$. $\alpha'_{ix}$ (or $\alpha'_{iz}$) is the polarizability derivative with respect to the layer $i$'s displacement along the $x$ (or $z$) direction; and $\Delta r_{ix}$ (or $\Delta r_{iz}$) is the layer $i$'s displacement along the $x$ (or $z$) direction in the interlayer vibration. For the shear vibration along the $x$ direction, the polarizability change is $\Delta \alpha = \sum_i \alpha'_{ix}\Delta r_{ix}$. As $\Delta \alpha$ and $\alpha'_{ix}$ are second-rank tensors, it equals to $\Delta \alpha_{\mu\nu} = \sum_i \alpha'_{ix,\mu\nu}\Delta r_{ix}$. Comparing this equation with Eq. 5, we find

$$\alpha'_{ix,\mu\nu} = -\sum_{B} \left\{ \hat{R}_{ix,B} \left[ \frac{\alpha'_{\parallel,B} + 2\alpha'_{\perp,B}}{3} \delta_{\mu\nu} + \left(\alpha'_{\parallel,B} - \alpha'_{\perp,B}\right) \left(\hat{R}_{i\mu,B}\hat{R}_{i\nu,B} - \frac{1}{3}\delta_{\mu\nu}\right) \right] \right.$$
$$\left. + \frac{\alpha_{\parallel,B} - \alpha_{\perp,B}}{R_{i,B}} \left[\hat{R}_{i\mu,B}\delta_{\nu x} + \hat{R}_{i\nu,B}\delta_{\mu x} - 2\hat{R}_{i\mu,B}\hat{R}_{i\nu,B}\hat{R}_{ix,B}\right] \right\}, \quad (6)$$

which suggests that $\alpha'_{ix}$ can be determined by the interlayer bond (length and direction), and interlayer bond polarizabilities.

According to Eq. 1, for the commonly used parallel polarization set-up in the backscattering geometry $z(xx)\bar{z}$, Raman intensity is proportional to $|\Delta \alpha_{xx}|^2$, and thus only the $xx$ components of the tensors need to be considered (i.e., $\mu = \nu = x$). Consequently, we have

$$\alpha'_{ix,xx} = -\sum_B \left\{ \frac{\alpha'_{\parallel,B} + 2\alpha'_{\perp,B}}{3} + (\alpha'_{\parallel,B} - \alpha'_{\perp,B})\hat{R}^2_{ix,B} - \frac{\alpha'_{\parallel,B} - \alpha'_{\perp,B}}{3} + 2\frac{\alpha_{\parallel,B} - \alpha_{\perp,B}}{R_{i,B}} - 2\frac{\alpha_{\parallel,B} - \alpha_{\perp,B}}{R_{i,B}}\hat{R}^2_{ix,B} \right\} \hat{R}_{ix,B}$$

$$= \sum_B C_{i,B}\hat{R}_{ix,B}, \tag{7}$$

where the coefficients $C_{i,B}$ are related to the properties of the interlayer bond $B$ connecting layer $i$ to a neighboring layer $i'$, such as the interlayer bond length and its $x$ component, and the interlayer bond polarizabilities and their radial derivatives. The change of the polarizability by the shear vibration is then

$$\Delta\alpha_{xx} = \sum_i \alpha'_{ix,xx}\Delta r_{ix} \tag{8}$$

For the breathing vibration along the $z$ direction, the change of the polarizability is $\Delta\alpha_{\mu\nu} = \sum_i \alpha'_{iz,\mu\nu}\Delta r_{iz}$, which is simplified as

$$\Delta\alpha_{xx} = \sum_i \alpha'_{iz,xx}\Delta r_{iz} \tag{9}$$

under the $z(xx)\bar{z}$ configuration. Similarly it can be shown that (see SI for details)

$$\alpha'_{iz,xx} = \sum_B C^*_{i,B}\hat{R}_{iz,B}, \tag{10}$$

where the coefficients $C^*_{i,B}$ are also related to the properties of the interlayer bond $B$.

Note that if every layer moves in the same manner along the $x$ direction (i.e., $\Delta r_{ix} = \Delta x$ while $\Delta r_{iy} = \Delta r_{iz} = 0$ for any layer $i$), the polarizability change by such acoustic vibration is $\Delta\alpha = (\sum_i \alpha'_{ix})\Delta x$. Such motion actually corresponds to the translation of the whole system by $\Delta x$, and the translational invariance of the system's polarizability requires $\Delta\alpha = 0$, which imposes that $\sum_i \alpha'_{ix} = 0$. Similarly by translating the system along the $y$ or $z$ direction, we can obtain $\sum_i \alpha'_{iy} = 0$ and $\sum_i \alpha'_{iz} = 0$. For the $xx$ components, we naturally have $\sum_i \alpha'_{ix,xx} = 0$ and $\sum_i \alpha'_{iz,xx} = 0$.

Meanwhile, the linear chain model, which also treats each layer as a rigid ball and the interlayer coupling as a spring, can provide the frequency and layer displacements of each interlayer mode for layered materials at any thickness.[9,30,34,39] It has been widely used to explain the thickness dependence of the LF Raman modes' frequencies and understand the interlayer coupling strength. For $N$-layer isotropic layered materials, such as graphene and $MoS_2$, there are $N-1$ doubly degenerate shear (S) modes and $N-1$ breathing (B) modes, and their frequencies are given by the

linear chain model as

$$\omega(S_j) = \omega(S_{\text{bulk}}) \sin\left(\frac{N-j}{2N}\pi\right)$$
$$\omega(B_j) = \omega(B_{\text{bulk}}) \sin\left(\frac{N-j}{2N}\pi\right), \qquad (11)$$

where $j = 1, 2, \ldots, N-2, N-1$ is the phonon branch index, $\omega(S_{\text{bulk}}) = (1/\pi c)\sqrt{K^\parallel/\mu}$, $\omega(B_{\text{bulk}}) = (1/\pi c)\sqrt{K^\perp/\mu}$, $K^\parallel$ ($K^\perp$) is the in-plane (out-of-plane) interlayer force constant per unit area, $\mu$ is the total mass per unit area of each layer, and $c$ is the speed of light.[9] Here $S_1$ ($B_1$) is the highest-frequency S (B) mode, while $S_{N-1}$ ($B_{N-1}$) is the lowest-frequency S (B) mode. For the $j$-th mode $S_j$ and $B_j$, the eigen-displacement of layer $i$ is[34]

$$\Delta r_{ix}(S_j) \propto \cos\left[\frac{(N-j)(2i-1)}{2N}\pi\right]$$
$$\Delta r_{iz}(B_j) \propto \cos\left[\frac{(N-j)(2i-1)}{2N}\pi\right]. \qquad (12)$$

Note that for graphene and $MX_2$, the interlayer force constants between different stacking polytypes considered in Fig. 1 are roughly the same, according to experimental frequencies and first-principles calculations.[47,49] Therefore, the stacking effects on the frequencies and layer displacements in Eq. 11 and Eq. 12 are ignored in this work, as our focus is the influence of stacking on Raman intensities.

In short, according to the interlayer bond polarizability model proposed here, $\alpha'_{ix,xx}$ in Eq. 7 and $\alpha'_{iz,xx}$ in Eq. 10 are related to the interlayer bond vectors in a simple fashion, and thus they can be obtained by determining the interlayer bond vectors of the system. Combined with the layer eigen-displacements in Eq. 12, the change of the polarizability $\Delta\alpha_{xx}$ is obtained for the $S_j$ mode based on Eq. 8 and for the $B_j$ mode based on Eq. 9, which subsequently yields the Raman intensities since $I \propto \frac{n_j+1}{\omega_j}|\Delta\alpha_{xx}|^2$. With their frequencies in Eq. 11, we can eventually obtain the LF Raman spectra after introducing Lorentzian broadening at room temperature ($T = 300$K).

## III. RESULTS AND DISCUSSIONS

### A. Stacking dependence of LF Raman intensities in graphene

As discussed before, AB and ABC are two common stacking types in multilayer graphene. The local interlayer stacking always assumes the pattern of AB, and consequently the interlayer bond length $R_{i,B}$, the interlayer bond polarizabilities and their radial derivatives can be treated as constants.[51] It follows that the coefficients $C_{i,B}$ and $C^*_{i,B}$ are also constants in multilayer graphene. In short, Eq. 7 and Eq. 10 can be simplified as $\alpha'_{ix,xx} = C \sum_B \hat{R}_{ix,B}$ and $\alpha'_{iz,xx} = C^* \sum_B \hat{R}_{iz,B}$, respectively. Starting from trilayer graphene, it has two stacking sequences: ABA and ABC. According to the schematics of the interlayer bonds shown in Fig. 2a for ABA stacking, layer 1 only has an interlayer bond with layer 2: the normalized bond vector is $\hat{R}_{1,2} = (\sin\theta, 0, \cos\theta)$; layer 2 has an interlayer bond with layer 1 and one with layer 3: the normalized bond vectors are $\hat{R}_{2,1} = (-\sin\theta, 0, -\cos\theta)$ and $\hat{R}_{2,3} = (-\sin\theta, 0, \cos\theta)$, respectively; layer 3 only has an interlayer bond with layer 2: the normalized bond vector is $\hat{R}_{3,2} = (\sin\theta, 0, -\cos\theta)$. Note that for layer $i$ and its neighboring layer $j$, there is a general relation $\hat{R}_{i,j} = -\hat{R}_{j,i}$. Therefore, for each layer in ABA stacking, we have

$$\begin{aligned}
\alpha'_{1x,xx} &= C\hat{R}_{1x,2} = C\sin\theta = \beta \\
\alpha'_{2x,xx} &= C(\hat{R}_{2x,1} + \hat{R}_{2x,3}) = -2C\sin\theta = -2\beta \\
\alpha'_{3x,xx} &= C\hat{R}_{3x,2} = C\sin\theta = \beta.
\end{aligned} \qquad (13)$$

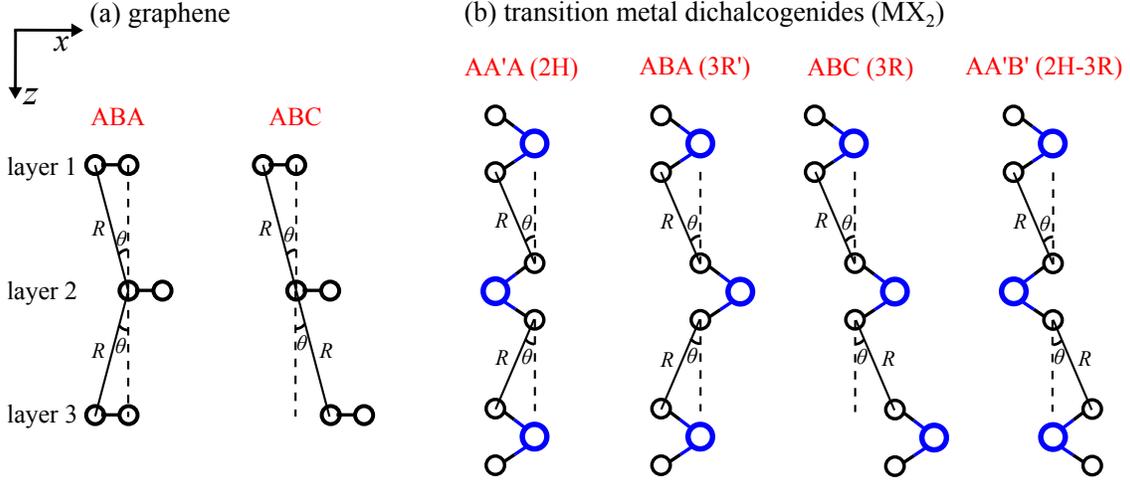

FIG. 2. (Color online) Schematic of the interlayer bonds in (a) trilayer graphene and (b) MX$_2$ in various stacking patterns.

On the other hand, for ABC stacking, the normalized bond vector for layer 1 is $\hat{R}_{1,2} = (\sin\theta, 0, \cos\theta)$; the normalized bond vectors for layer 2 are $\hat{R}_{2,1} = (-\sin\theta, 0, -\cos\theta)$ and $\hat{R}_{2,3} = (\sin\theta, 0, \cos\theta)$, respectively; the normalized bond vector for layer 3 is $\hat{R}_{3,2} = (-\sin\theta, 0, -\cos\theta)$. Therefore, for each layer in ABC stacking, we have

$$\alpha'_{1x,xx} = C\hat{R}_{1x,2} = C\sin\theta = \beta$$
$$\alpha'_{2x,xx} = C(\hat{R}_{2x,1} + \hat{R}_{2x,3}) = 0$$
$$\alpha'_{3x,xx} = C\hat{R}_{3x,2} = -C\sin\theta = -\beta. \quad (14)$$

Interestingly, $\alpha'_{1x,xx} = \alpha'_{3x,xx}$ for ABA stacking as layer 1 and layer 3 are related by mirror symmetry, while $\alpha'_{1x,xx} = -\alpha'_{3x,xx}$ as layer 1 and layer 3 are related by inversion symmetry.[47] In either stacking, we have $\alpha'_{1x,xx} + \alpha'_{2x,xx} + \alpha'_{3x,xx} = 0$, which corresponds to the general relation discussed above. In short, because the $x$ components of the normalized bond vectors are different between two stackings, $(\alpha'_{1x,xx}, \alpha'_{2x,xx}, \alpha'_{3x,xx})$ are $(\beta, -2\beta, \beta)$ for ABA stacking and $(\beta, 0, -\beta)$ for ABC stacking.[51] Injecting this information into Eq. 8, the polarizability change by the shear vibrations is $\Delta\alpha_{xx} = \beta(\Delta r_{1x} - 2\Delta r_{2x} + \Delta r_{3x})$ for ABA stacking, and $\Delta\alpha_{xx} = \beta(\Delta r_{1x} - \Delta r_{3x})$ for ABC stacking. Furthermore, there are two shear modes ($S_2$ and $S_1$) for trilayer graphene. According to Eq. 12 for both stackings, the normalized layer displacements $(\Delta r_{1x}, \Delta r_{2x}, \Delta r_{3x})$ are $\frac{1}{\sqrt{2}}(1, 0, -1)$ for the $S_2$ mode and $\frac{1}{\sqrt{1.5}}(0.5, -1, 0.5)$ for the $S_1$ mode (Fig. 3a). It follows that

$$\Delta\alpha_{xx}(\text{ABA}, S_2) = 0; \qquad \Delta\alpha_{xx}(\text{ABA}, S_1) = \sqrt{6}\beta;$$
$$\Delta\alpha_{xx}(\text{ABC}, S_2) = \sqrt{2}\beta; \qquad \Delta\alpha_{xx}(\text{ABC}, S_1) = 0.$$

Since the Raman intensity $I$ is proportional to $|\Delta\alpha_{xx}|^2$ (see Eq. 1), these results suggest that the $S_1$ ($S_2$) mode should be observed exclusively in the Raman scattering of ABA (ABC) stacking sequence, as shown in Fig. 3b. The calculated Raman spectra based on our interlayer bond polarizability model agree with the experimental data by Lui et al.[47]

Turning to the $z$ direction, for each layer in ABA stacking, we have

$$\alpha'_{1z,xx} = C^*\hat{R}_{1z,2} = C^*\cos\theta = \gamma$$
$$\alpha'_{2z,xx} = C^*(\hat{R}_{2z,1} + \hat{R}_{2z,3}) = 0$$
$$\alpha'_{3z,xx} = C^*\hat{R}_{3z,2} = -C^*\cos\theta = -\gamma. \quad (15)$$

Conversely, for each layer in ABC stacking, we have

$$\alpha'_{1z,xx} = C^*\hat{R}_{1z,2} = C^*\cos\theta = \gamma$$
$$\alpha'_{2z,xx} = C^*(\hat{R}_{2z,1} + \hat{R}_{2z,3}) = 0$$
$$\alpha'_{3z,xx} = C^*\hat{R}_{3z,2} = -C^*\cos\theta = -\gamma. \quad (16)$$

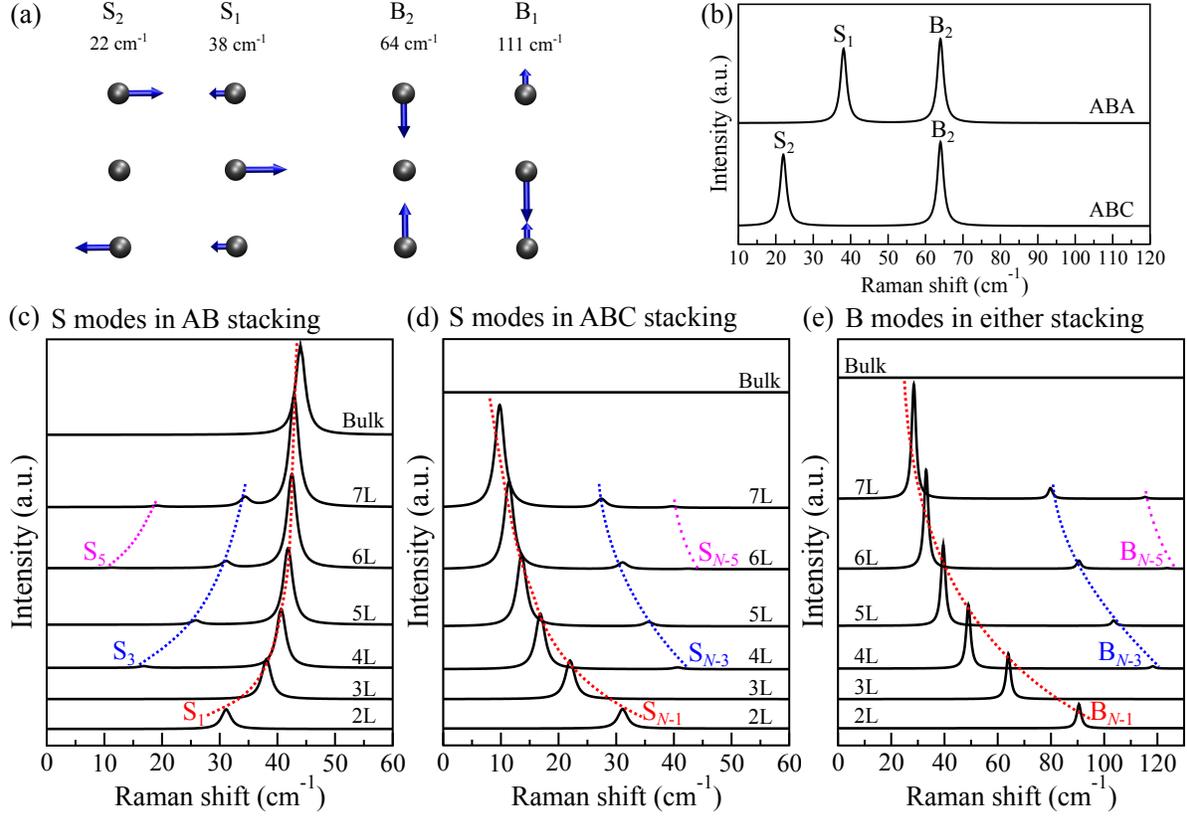

FIG. 3. (Color online) (a) Interlayer vibrations of trilayer graphene, where each layer is simplified as a single object. The arrows indicate both magnitude and direction of the vibrations. (b) Calculated LF Raman spectra for ABA- and ABC-stacked trilayer graphene. For graphene, calculated thickness dependence of Raman scattering of (c) S modes in AB stacking, (d) S modes in ABC stacking, and (e) B modes in AB or ABC stacking. Dashed lines indicate the frequency evolution trends of the modes.

The obvious difference from the $x$ direction is that $\alpha'_{1z,xx} = -\alpha'_{3z,xx} = \gamma$, and $\alpha'_{2z,xx} = 0$ in both ABA and ABC stackings, since the $z$ components of the interlayer bond vectors are stacking independent. Again, we verify that $\alpha'_{1z,xx} + \alpha'_{2z,xx} + \alpha'_{3z,xx} = 0$. Adding this information into Eq. 9, the polarizability change by the breathing vibrations is $\Delta \alpha_{xx} = \gamma(\Delta r_{1z} - \Delta r_{3z})$ for both stacking configurations. Furthermore, there are two breathing modes ($B_2$ and $B_1$) for trilayer graphene. According to Eq. 12 for both stackings, the normalized layer displacements ($\Delta r_{1z}, \Delta r_{2z}, \Delta r_{3z}$) are $\frac{1}{\sqrt{2}}(1,0,-1)$ for the B2 mode and $\frac{1}{\sqrt{1.5}}(0.5,-1,0.5)$ for the B1 mode (Fig. 3a), and we find

$$\Delta \alpha_{xx}(\text{ABA or ABC}, B_2) = \sqrt{2}\gamma;$$
$$\Delta \alpha_{xx}(\text{ABA or ABC}, B_1) = 0.$$

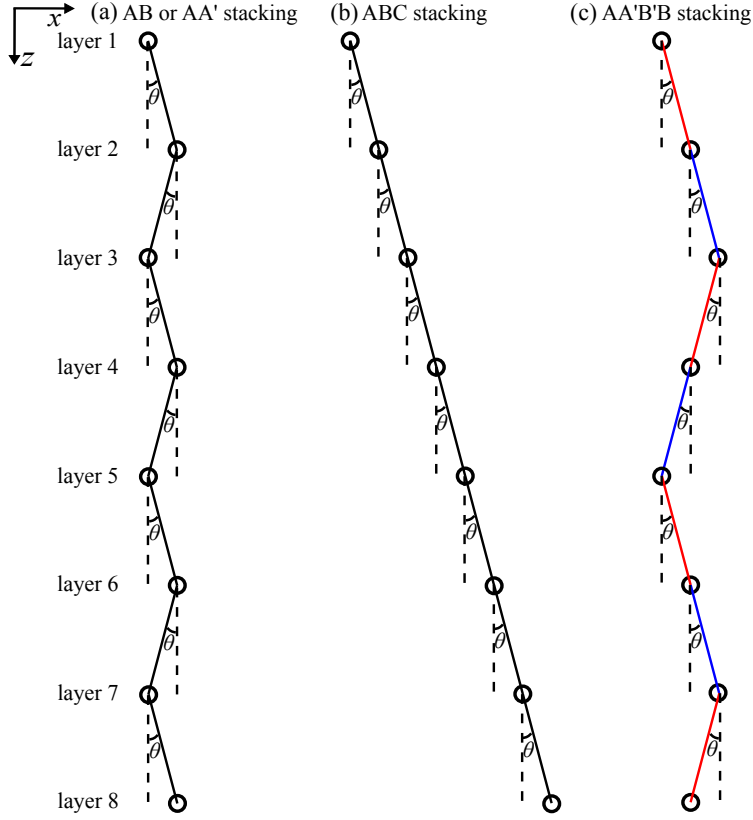

FIG. 4. (Color online) Schematic of the interlayer bonds in $N$-layer for (a) AB or AA′ stacking, (b) ABC stacking, and (c) AA′B′B stacking types. Each layer is simplified as a single object for interlayer vibrations. AB or AA′ stacking repeats every two layer, ABC stacking repeats every three layers, and AA′B′B stacking repeats every four layers. In (c) AA′B′B stacking, the interlayer bonds corresponding to AA′ and AB stackings are differentiated by red and blue colors, respectively.

This result explains why the $B_2$ mode is observed in both stackings while the $B_1$ mode is not (Fig. 3b), in agreement with prior experimental work.[47] Note that to compute LF Raman spectra in Fig. 3b, $\gamma \approx 3\beta$ was assumed to fit with the experimental spectra in Ref. 47.

Moving to $N$-layer graphene, with the exception of layer 1 and layer $N$ which have only one interlayer bond, other interior layers have two interlayer bonds. For an interior layer $i \neq 1, N$, the $x$ components of these two normalized bond vectors obey a relation $\hat{R}_{ix,i-1} = \hat{R}_{ix,i+1}$ in AB stacking, while $\hat{R}_{ix,i-1} = -\hat{R}_{ix,i+1}$ in ABC stacking, as illustrated in Fig. 4. Such contrast gives rise to dramatically different $\alpha'_{ix,xx}$ between two stackings. Specifically, for AB stacking, $\alpha'_{1x,xx} = \beta$, $\alpha'_{Nx,xx} = \beta$ for odd $N$ or $\alpha'_{Nx,xx} = -\beta$ for even $N$, and $\alpha'_{2x,xx} = -2\beta, \alpha'_{3x,xx} = 2\beta, \alpha'_{4x,xx} = -2\beta, \alpha'_{5x,xx} = 2\beta, \ldots$, where there is a repeated pattern of $-2\beta, 2\beta$ for the interior layers. For ABC stacking, the situaion it is much simpler: $\alpha'_{1x,xx} = \beta$, $\alpha'_{Nx,xx} = -\beta$, while for all interior layers, $\alpha'_{ix,xx} = C(\hat{R}_{ix,i-1} + \hat{R}_{ix,i+1}) = 0$ (more details in Section S2 in SI). Incorporating this information into Eq. 8, the polarizability change of $N$-layer graphene by the shear vibrations is

$$\Delta\alpha_{xx}(\text{AB}) = \beta(\Delta r_{Nx} - \Delta r_{1x}) + 2\beta \sum_{i=1,3,5}^{m} \left(\Delta r_{ix} - \Delta r_{(i+1)x}\right);$$

$$\Delta\alpha_{xx}(\text{ABC}) = \beta(\Delta r_{1x} - \Delta r_{Nx}), \tag{17}$$

where $m$ is the largest odd number smaller than $N$ (i.e., $m = N - 2$ for odd $N$, while $m = N - 1$ for even $N$). The normalized layer displacements $\Delta r_{ix}$ are given by Eq. 12, and the frequencies of all $N - 1$ S modes in $N$-layer graphene are given by Eq. 11, where the frequency of the S mode in bulk graphite[34] is $\omega(\text{S}_{\text{bulk}}) \approx 44.0$ cm$^{-1}$. With Raman intensity $I \propto \dfrac{n_j + 1}{\omega_j}|\Delta\alpha_{xx}|^2$, we can obtain the intensities of the S modes in both stackings as shown in Fig. 3. For the $N - 1$ S modes in $N$-layer graphene ($S_1$, $S_2$, $S_3$, ..., $S_{N-1}$ with the ordering going from the highest to lowest frequency), in AB stacking, starting from the highest-frequency one, only $S_1$, $S_3$, $S_5$, ... can be observed with an intensity trend $I(S_1) > I(S_3) > I(S_5) > \ldots$ (Fig. 3c); in ABC stacking, the trend is the opposite and starting from the lowest-frequency one, only $S_{N-1}$, $S_{N-3}$, $S_{N-5}$, ... can be observed with an intensity trend $I(S_{N-1}) > I(S_{N-3}) > I(S_{N-5}) > \ldots$ (Fig. 3d). In other words, the observable S modes in AB stacking include the highest-frequency branch ($S_1$), third highest-frequency branch ($S_3$), fifth highest-frequency branch ($S_5$), etc, and their frequencies increase with increasing thickness according to the linear chain model (Fig. 3c); on the contrary, the observable S modes in ABC stacking include the lowest-frequency branch ($S_{N-1}$), third lowest-frequency branch ($S_{N-3}$), fifth lowest-frequency branch ($S_{N-5}$), etc, and their frequencies decrease with increasing thickness (Fig. 3d). These results from the interlayer bond polarizability model are consistent with the first-principles calculations by Luo et al.[51] Such opposite trends between AB and ABC stackings underscore that the S modes' intensities can facilitate the stacking identification of multilayer graphene. Taking 6L as an example in Figs. 3c and 3d, we have $I(S_1) > I(S_3) > I(S_5)$ for AB stacking, while $I(S_5) > I(S_3) > I(S_1)$ for ABC stacking.

We now carry out further analysis to understand the distinct Raman response of the S modes to stacking. For AB stacking, $\Delta\alpha_{xx}$ is dominated by the second term in Eq. 17, so its largest value occurs to the S mode that has every adjacent layers vibrating in opposite directions. According to Eq. 12 and vibration schematics shown in Fig. S3, the $S_1$ mode (the highest-frequency S branch) satisfies such condition and thus exhibits the largest intensity in $N$-layer samples, while other observable modes like $S_3$ and $S_5$ have relatively lower intensities, as shown in Fig. 3c. For modes like $S_2$ and $S_4$, the polarizability change by each layer's displacement cancels each other, yielding $\Delta\alpha_{xx} = 0$ and subsequently zero intensities. For the bulk in AB stacking, there are no exterior layers due to the periodic boundary conditions, and thus $\alpha'_{1x,xx} = 2\beta$, $\alpha'_{2x,xx} = -2\beta$, $\alpha'_{3x,xx} = 2\beta$, $\alpha'_{4x,xx} = -2\beta$, ... where there is a repeated pattern of $2\beta, -2\beta$ for all layers. It follows that

$$\Delta\alpha_{xx}(\text{AB, bulk}) = 2\beta \sum_{i=1,3,5,...} \left(\Delta r_{ix} - \Delta r_{(i+1)x}\right),$$

which is the limit of Eq. 17 when $N \to \infty$. The S mode in the bulk also has every adjacent layers vibrating in the opposite directions. These results justify why the $S_1$ mode in $N$-layer is called the bulk-like mode, and its frequency and intensity approach those of the bulk S mode when $N \to \infty$ (Fig. 3c). As for ABC stacking, Eq. 17 shows that $\Delta\alpha_{xx}$ is only related to the displacements of the top and bottom layers, and the largest value occurs to the S mode that has the largest opposite displacements of layer 1 and layer $N$. According to Eq. 12 and Fig. S3, $S_{N-1}$ (the lowest-frequency S branch) satisfies such condition and thus exhibits the largest intensity in $N$-layer, while other observable modes like $S_{N-3}$ and $S_{N-5}$ have relatively lower intensities due to smaller opposite displacements of layer 1 and layer $N$, as shown in Fig. 3d. For modes like $S_{N-2}$ and $S_{N-4}$, the same displacements between the top and bottom layers result in no polarizability changes and thus zero intensities. For the bulk in ABC stacking, we always have $\hat{R}_{ix,i-1} = -\hat{R}_{ix,i+1}$ and hence $\alpha'_{ix,xx} = C(\hat{R}_{ix,i-1} + \hat{R}_{ix,i+1}) = 0$ for any layer $i$. Consequently any shear vibration does not change the polarizability, and the intensities of the S modes are zero for the ABC-stacked bulk (Fig. 3d).

On the other hand, for an interior layer $i$, the $z$ components of the two normalized interlayer bond vectors always assume a relation $\hat{R}_{iz,i-1} = -\hat{R}_{iz,i+1}$ regardless of the stacking type. Subsequently, $\alpha'_{1z,xx} = \gamma$, $\alpha'_{Nz,xx} = -\gamma$, while for all interior layers, $\alpha'_{iz,xx} = C^*(\hat{R}_{iz,i-1} + \hat{R}_{iz,i+1}) = 0$ (more details in Section S2 in SI). These results are very similar to those obtained for the S modes in ABC stacking. Adding into Eq. 9, the polarizability change of $N$-layer graphene by the breathing vibrations is simply

$$\Delta\alpha_{xx} = \gamma(\Delta r_{1z} - \Delta r_{Nz}) \tag{18}$$

for both AB and ABC stackings. Similarly, the normalized layer displacements $\Delta r_{iz}$ are given by Eq. 12, and the frequencies of all $N-1$ B modes in $N$-layer graphene are given by Eq. 11, where the frequency of the B mode in bulk graphite[35,74] is $\omega(B_{\text{bulk}}) \approx 128.0$ cm$^{-1}$. The intensities of the B modes can be subsequently obtained, which are the same in both stackings (Fig. 3e). Eq. 18 again shows that $\Delta\alpha_{xx}$ is only related to the displacements of the top and bottom layers, and the largest value occurs to the B mode that has the largest opposite displacements of layer 1 and layer $N$. According to Eq. 12 and Fig. S3, $B_{N-1}$ (the lowest-frequency B branch) satisfies such condition and thus exhibits the largest intensity in $N$-layer, as shown in Fig. 3e. Consequently, for the $N-1$ B modes in $N$-layer graphene in both AB and ABC stackings, starting from the lowest-frequency one, only $B_{N-1}$, $B_{N-3}$, $B_{N-5}$, ... can be observed with an intensity trend $I(B_{N-1}) > I(B_{N-3}) > I(B_{N-5}) > \ldots$ (Fig. 3e). Such trend is similar to the S modes in ABC stacking. Again, the observable B modes in both stackings include the lowest-frequency branch ($B_{N-1}$), third lowest-frequency branch ($B_{N-3}$), fifth lowest-frequency branch ($B_{N-5}$), etc, and their frequencies decrease with increasing thickness (Fig. 3e). For the bulk in either AB or ABC stacking, we always have $\hat{R}_{iz,i-1} = -\hat{R}_{iz,i+1}$ and hence $\alpha'_{iz,xx} = C^*(\hat{R}_{iz,i-1} + \hat{R}_{iz,i+1}) = 0$ for any layer $i$. Consequently any breathing vibration does not change the polarizability, and the intensities of the B modes are zero for the bulk (Fig. 3e). This explains why the B modes cannot be observed in bulk graphite.

To summarize this section, for multilayer graphene, the $z$ (i.e., out-of-plane) components of the interlayer bond vectors do not change with the in-plane stacking variation, and thus $\alpha'_{iz,xx}$ are stacking independent, so the intensities of the B modes are stacking independent and cannot be used for stacking identification; in contrast, the $x$ (i.e., in-plane) components of the interlayer bond vectors can change with in-plane stacking variation, and thus $\alpha'_{ix,xx}$ can be highly stacking dependent, so the intensities of the S modes show unique stacking dependence for its identification. For instance, among the shear modes, the highest-frequency one ($S_1$) has the largest Raman intensity in AB-stacked multilayer graphene, while the lowest-frequency one ($S_{N-1}$) has the largest Raman intensity in ABC-stacked systems.[48,51] These findings about graphene are in fact very generalized, and can be also applied to many other 2D materials as discussed below.

### B. Stacking dependence of LF Raman intensities in MX$_2$

Compared to mono-elemental materials like graphene, the stacking patterns are significantly more complicated in trilayer MX$_2$ (M = Mo or W; X = S or Se). Unlike trilayer graphene where the interlayer stacking is always AB, there are two distinctively different interlayer stacking patterns in trilayer MX$_2$: AA$'$ (or 2H) and AB (or 3R), as shown in Fig. 2b. According to Eq. 7, for bilayer MX$_2$ in AA$'$ stacking, we have

$$\alpha'_{1x,xx} = C_{1,2}\hat{R}_{1x,2} = C(AA')\sin\theta = \beta_1$$
$$\alpha'_{2x,xx} = C_{2,1}\hat{R}_{2x,1} = -C(AA')\sin\theta = -\beta_1,$$

where $C_{1,2} = C_{2,1} = C(AA')$, the coefficient related to the interlayer bond polarizability and its derivatives in AA′ stacking. For the shear mode $S_1$, the normalized displacements $(\Delta r_{1x}, \Delta r_{2x})$ are $\frac{1}{\sqrt{2}}(1,-1)$, so adding into Eq. 8, we can obtain the polarizability change as $\Delta\alpha_{xx}(AA') = \sqrt{2}\beta_1$. Similarly, for bilayer $MX_2$ in AB stacking, we have

$$\alpha'_{1x,xx} = C_{1,2}\hat{R}_{1x,2} = C(AB)\sin\theta = \beta_2$$
$$\alpha'_{2x,xx} = C_{2,1}\hat{R}_{2x,1} = -C(AB)\sin\theta = -\beta_2,$$

where $C_{1,2} = C_{2,1} = C(AB)$, the coefficient related to the interlayer bond polarizability and its derivatives in AB stacking. The polarizability change by the $S_1$ mode is then $\Delta\alpha_{xx}(AB) = \sqrt{2}\beta_2$. Since the relative layer-layer atomic alignments are changed between AA′ and AB stackings, the interlayer bond polarizability and its derivatives (i.e., $\beta_1$ and $\beta_2$) are different. This is also reflected by the different LF Raman response of bilayer $MX_2$ in the two types of stacking.[49,66] Taking MoSe$_2$ as an example, Puretzky et al. found that the intensity of the $S_1$ mode in bilayer MoSe$_2$ drops from AA′ (2H) to AB (3R) stacking by a factor of 5.4, and such intensity drop was also corroborated by first-principles Raman calculations.[49] Because the frequency of the $S_1$ mode barely changes from AA′ to AB stacking, we simply have the intensity ratio $\frac{I(AA')}{I(AB)} = \frac{|\Delta\alpha_{xx}(AA')|^2}{|\Delta\alpha_{xx}(AB)|} = \frac{|\beta_1|^2}{|\beta_2|}$. Subsequently, we find that the magnitude ratio $|\beta_1|/|\beta_2| = \sqrt{5.4} = 2.32$, and the corresponding stacking-dependent Raman scattering of bilayer MoSe$_2$ is shown in Fig. 5a.

Moving to trilayer $MX_2$, AA′A and ABA stacking sequences have similar normalized interlayer bond vectors to ABA stacking in trilayer graphene, as shown in Fig. 2. Following the similar procedures in Eq. 13, for trilayer $MX_2$ in AA′A stacking, we find

$$\alpha'_{1x,xx} = C_{1,2}\hat{R}_{1x,2} = C(AA')\sin\theta = \beta_1$$
$$\alpha'_{2x,xx} = C_{2,1}\hat{R}_{2x,1} + C_{2,3}\hat{R}_{2x,3} = -2C(AA')\sin\theta = -2\beta_1 \quad (19)$$
$$\alpha'_{3x,xx} = C_{3,2}\hat{R}_{3x,2} = C(AA')\sin\theta = \beta_1,$$

where $C_{1,2} = C_{2,1} = C_{2,3} = C_{3,2} = C(AA')$. Similarly, in ABA stacking, we have

$$\alpha'_{1x,xx} = C(AB)\sin\theta = \beta_2$$
$$\alpha'_{2x,xx} = -2C(AB)\sin\theta = -2\beta_2$$
$$\alpha'_{3x,xx} = C(AB)\sin\theta = \beta_2. \quad (20)$$

Note that although the forms of $\alpha'_{ix,xx}$ here are the same as those in Eq. 13 for trilayer graphene in ABA stacking, the coefficients $C(AA')$ and $C(AB)$ are different. On the other hand, the other two stacking sequences in trilayer MX$_2$, ABC and AA′B′ have similar normalized interlayer bond vectors compared to ABC stacking in trilayer graphene (Fig. 2). Following a procedure similar to that used to derive Eq. 14, for trilayer MX$_2$ in ABC stacking, we have

$$\begin{aligned}
\alpha'_{1x,xx} &= C_{1,2}\hat{R}_{1x,2} = C(\mathrm{AB})\sin\theta = \beta_2 \\
\alpha'_{2x,xx} &= C_{2,1}\hat{R}_{2x,1} + C_{2,3}\hat{R}_{2x,3} = 0 \\
\alpha'_{3x,xx} &= C_{3,2}\hat{R}_{3x,2} = -C(\mathrm{AB})\sin\theta = -\beta_2,
\end{aligned} \qquad (21)$$

where $C_{1,2} = C_{2,1} = C_{2,3} = C_{3,2} = C(\mathrm{AB})$. The forms of $\alpha'_{ix,xx}$ here are the same as those in Eq. 14 for trilayer graphene in ABC stacking. At AA′B′ stacking, the situation is more complicated owing to the mixture of AA′ and AB stackings (A′B′ stacking equivalent to AB):

$$\begin{aligned}
\alpha'_{1x,xx} &= C_{1,2}\hat{R}_{1x,2} = C(\text{AA}')\sin\theta = \beta_1 \\
\alpha'_{2x,xx} &= C_{2,1}\hat{R}_{2x,1} + C_{2,3}\hat{R}_{2x,3} \\
&= -C(\text{AA}')\sin\theta + C(\text{AB})\sin\theta = -\beta_1 + \beta_2 \\
\alpha'_{3x,xx} &= C_{3,2}\hat{R}_{3x,2} = -C(\text{AB})\sin\theta = -\beta_2,
\end{aligned} \quad (22)$$

where $C_{1,2} = C_{2,1} = C(\text{AA}')$ while $C_{2,3} = C_{3,2} = C(\text{AB})$. In all four stacking configurations found in trilayer MX$_2$, we still have $\alpha'_{1x,xx} + \alpha'_{2x,xx} + \alpha'_{3x,xx} = 0$, the general relation discussed above. Like trilayer graphene, there are two shear modes (S$_2$ and S$_1$) for trilayer MX$_2$, and the normalized layer displacements $(\Delta r_{1x}, \Delta r_{2x}, \Delta r_{3x})$ are $\frac{1}{\sqrt{2}}(1,0,-1)$ for the S$_2$ mode and $\frac{1}{\sqrt{1.5}}(0.5,-1,0.5)$ for the S$_1$ mode (Fig. 3a). With $\Delta\alpha_{xx} = \sum_{i=1}^{3} \alpha'_{ix,xx}\Delta r_{ix}$, the polarizability change by the shear vibrations can be derived as follows:

$$\begin{aligned}
&\Delta\alpha_{xx}(\text{AA}'\text{A}, \text{S}_2) = 0; &&\Delta\alpha_{xx}(\text{AA}'\text{A}, \text{S}_1) = \sqrt{6}\beta_1; \\
&\Delta\alpha_{xx}(\text{ABA}, \text{S}_2) = 0; &&\Delta\alpha_{xx}(\text{ABA}, \text{S}_1) = \sqrt{6}\beta_2; \\
&\Delta\alpha_{xx}(\text{ABC}, \text{S}_2) = \sqrt{2}\beta_2; &&\Delta\alpha_{xx}(\text{ABC}, \text{S}_1) = 0; \\
&\Delta\alpha_{xx}(\text{AA}'\text{B}', \text{S}_2) = \frac{\beta_1 + \beta_2}{\sqrt{2}}; &&\Delta\alpha_{xx}(\text{AA}'\text{B}', \text{S}_1) = \sqrt{1.5}(\beta_1 - \beta_2).
\end{aligned}$$

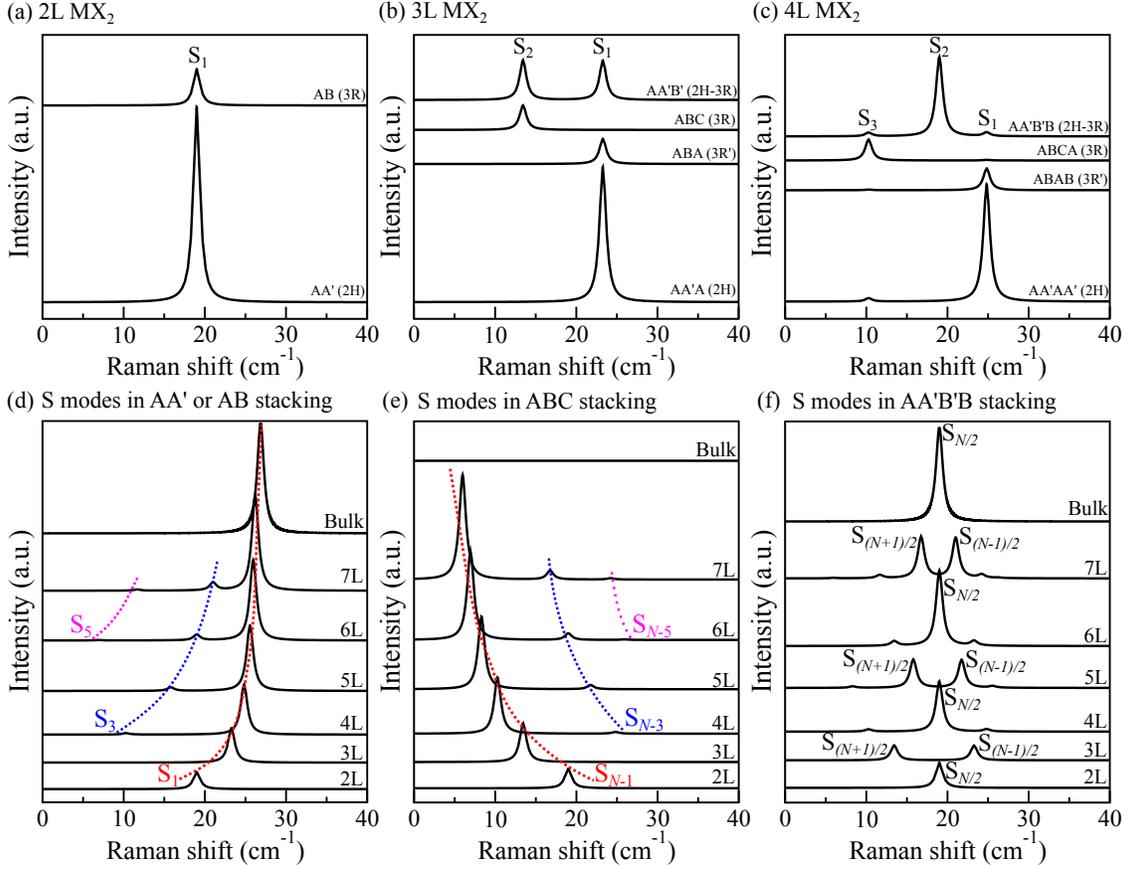

FIG. 5. (Color online) Calculated Raman spectra of S modes in (a) bilayer, (b) trilayer, and (c) four-layer MoSe$_2$ in different stacking configurations. Calculated thickness dependence of S modes in (d) AA′ (2H) or AB (3R′) stacking, (e) ABC (3R) stacking, and (f) AA′B′B (2H-3R mixed) stacking. Similar results apply to other group-6 MX$_2$. Dashed lines indicate the frequency evolution trends of the modes. In (d), the S modes in AB stacking exhibit systemically lower intensities compared to those in AA′ stacking, though the trends are the same.

For both AA′A (2H) and ABA (3R′) stackings, the S$_2$ peak intensity is zero while only the S$_1$ mode can be observed, similar to ABA stacking in 3L graphene. In addition, the fact of $|\beta_1| > |\beta_2|$ indicates that the S$_1$ peak intensity of AA′A stacking is higher than that of ABA stacking (Fig. 5b), thereby enabling their differentiation as well. In contrast, for ABC (3R) stacking, the S$_1$ peak intensity is zero while only the S$_2$ mode can be observed, similar to ABC stacking in 3L graphene. What is unique in trilayer MX$_2$ occurs in AA′B′ stacking (the mixture of AA′ and AB or mixture of 2H and 3R), where both S$_2$ and S$_1$ modes have non-zero intensities. Such stacking-dependent LF Raman response obtained from our interlayer bond polarizability model (Fig. 5b) can explain existing experimental data for 3L MoSe$_2$[49,50,52] and 3L MoS$_2$,[53] demonstrating that the S modes' intensities can be stacking fingerprints of MX$_2$.

Note that for 3L MoSe$_2$ in AA′B′ stacking, Puretzky et al. reported that the S$_2$ and S$_1$ modes exhibit nearly equal intensities.[49] Interestingly, if $\beta_1$ and $\beta_2$ are assumed to be real variables (no imaginary parts) as in the common non-resonant Raman modeling, we cannot obtain $I(\text{S}_2) = I(\text{S}_1)$ for AA′B′ stacking. In reality, the polarizability (or dielectric function) has both real and imaginary parts due to the light absorption in experimental resonant Raman scattering.[75,76] Thus $\beta_1$ and $\beta_2$ are complex variables: $\beta_1 = |\beta_1|e^{i\phi_1}; \beta_2 = |\beta_2|e^{i\phi_2}$, where $|\beta_1| = 2.32|\beta_2|$ obtained from bilayer MoSe$_2$, and $\phi_1$ and $\phi_2$ are the phase angles, respectively. For 3L MoSe$_2$ in AA′B′ stacking, in order to have $I(S_2) = I(S_1)$, it is required that $|\phi_1 - \phi_2| \approx 88.74°$ (detailed derivations in Section S3 in SI). Here we assume $|\beta_1| = 2.32$ and $\phi_1 = 118.74°$, while $|\beta_2| = 1.00$ and $\phi_2 = 30.00°$ without loss of generality. These parameters give rise to nearly equal intensities between the $S_2$ and $S_1$ modes for 3L MoSe$_2$ in AA′B′ stacking (Fig. 5b).

Moving to $N$-layer MX$_2$, once again AA′ and AB stacking types have similar interlayer bond vectors to AB stacking in $N$-layer graphene (Fig. 4). Following the same procedure used for graphene, we also have $\alpha'_{1x,xx} = \beta$, $\alpha'_{Nx,xx} = \beta$ for odd $N$ or $\alpha'_{Nx,xx} = -\beta$ for even $N$, and $\alpha'_{2x,xx} = -2\beta, \alpha'_{3x,xx} = 2\beta, \alpha'_{4x,xx} = -2\beta, \alpha'_{5x,xx} = 2\beta, \ldots$, where there is a repeated pattern of $-2\beta, 2\beta$ for the interior layers. Here $\beta = \beta_1$ for AA′ stacking, while $\beta = \beta_2$ for AB stacking. In contrast, ABC stacking in $N$-layer MX$_2$ has similar interlayer bond vectors to ABC stacking in $N$-layer graphene, and hence similar to graphene: $\alpha'_{1x,xx} = \beta_2$, $\alpha'_{Nx,xx} = -\beta_2$, while for all interior layers, $\alpha'_{ix,xx} = 0$ (more details in Section S2 in SI). The AA′B′B stacking in $N$-layer MX$_2$ is more complicated due to the mixture of AA′ and AB stackings (Fig. 4c) and we find $\alpha'_{1x,xx} = \beta_1, \alpha'_{2x,xx} = -\beta_1 + \beta_2, \alpha'_{3x,xx} = -\beta_2 - \beta_1, \alpha'_{4x,xx} = \beta_1 - \beta_2, \alpha'_{5x,xx} = \beta_2 + \beta_1, \ldots$, where for an interior layer $i$, $\alpha'_{ix,xx} = -\alpha'_{(i+2)x,xx}$, and thus $\alpha'_{ix,xx} = \alpha'_{(i+4)x,xx}$, since AA′B′B stacking repeats every four layers (see Eq. S17 in Section S2 in SI). With the normalized layer displacements $\Delta r_{ix}$ given by Eq. 12 and the frequencies of all $N-1$ S modes given by Eq. 11, we can obtain the intensities of the S modes in all four stacking polytypes for $N$-layer MX$_2$.

Selecting MoSe$_2$ as an example without loss of generality, the frequency of its bulk S mode is $\omega(S_{\text{bulk}}) \approx 26.9$ cm$^{-1}$, and Raman spectra of the S modes are shown in Fig. 5. For AA′ (2H) or AB (3R′) stacking type in $N$-layer MoSe$_2$ samples that share similar interlayer bond vectors to AB stacking type in $N$-layer graphene, starting from the highest-frequency one, only $S_1$, $S_3$, $S_5, \ldots$ can be observed with an intensity trend $I(S_1) > I(S_3) > I(S_5) > \ldots$ (Fig. 5d). Such trend is the same as that of AB-stacked graphene shown in Fig. 3c. Because $\alpha'_{ix,xx}$ assumes a repeated pattern of $-2\beta, 2\beta$ for the interior layers, S modes with vibrations close to every adjacent layers moving in the opposite directions show larger signals (Eq. 17), and the $S_1$ mode (the highest-frequency S branch) best satisfies such condition and thus exhibits the largest intensity in $N$-layer MoSe$_2$. These results from our model in Fig. 5d explain why only the highest-frequency, third highest-frequency and fifth highest-frequency S modes ($S_1$, $S_3$ and $S_5$) can be observed for natural AA′-stacked (2H-stacked) MoS$_2$ and WSe$_2$ samples at different thicknesses.[36,39] Furthermore, the prediction of $I(S_1) > I(S_3) > I(S_5)$ is almost quantitatively consistent with experimental data.[36,39] Note that although $S_3$, $S_5$, $S_7$, ... are in principle observable starting from 4L, 6L, 8L, ..., respectively, experimentally they could be too weak to be observed at all, or could be strong enough to be observed only at thicker layers. This is particularly true for $S_5$, $S_7$, etc.[36,39] The bulk-like $S_1$ peak also approaches the bulk S peak when $N \to \infty$ (Fig. 5d), consistent with the experimental measurements.[36,39] Although natural samples generally assume AA′ (2H) stacking, AB stacking can be found in synthetic samples. The S modes in AB stacking exhibit systemically lower intensities than those in AA′ stacking (see Figs. 5a-c), which can help to differentiate the two stacking types in MoSe$_2$.

In contrast, for ABC (3R) stacking type in $N$-layer MoSe$_2$ that shares similar interlayer bond vectors to ABC stacking in $N$-layer graphene, starting from the lowest-frequency one, only $S_{N-1}$, $S_{N-3}$, $S_{N-5}$, ... can be observed with an intensity trend $I(S_{N-1}) > I(S_{N-3}) > I(S_{N-5}) > ...$ (Fig. 5e). Such trend is the same as that of ABC-stacked graphene shown in Fig. 3d. According to Eq. 17, the intensities of S modes in ABC stacking are only related to the displacement difference between the top and bottom layers, and thus $S_{N-1}$ (the lowest-frequency S branch) with the largest displacement difference exhibits the largest intensity in $N$-layer. For S modes like $S_{N-2}$ and $S_{N-4}$, the displacement difference between the top and bottom layers is zero (Fig. S3), and hence they have zero intensities. Similar to bulk graphene in ABC stacking, we always have $\hat{R}_{ix,i-1} = -\hat{R}_{ix,i+1}$ and hence $\alpha'_{ix,xx} = C(\hat{R}_{ix,i-1} + \hat{R}_{ix,i+1}) = 0$ for any layer $i$ in bulk MoSe$_2$ in ABC stacking. Consequently any shear vibration does not change the polarizability, and the intensities of the S modes are zero for the ABC-stacked bulk (Fig. 5e). It is clear that the Raman responses of S modes to AA′ (or AB) stacking and ABC stacking are opposite for any thickness (Figs. 5d and 5e). For AA′B′B (2H-3R) stacking type in $N$-layer MoSe$_2$, since it is the mixture of AA′ and AB stackings (i.e., mixture of 2H and 3R stackings), the S modes in the middle, instead of the highest-frequency or lowest-frequency ones, dominate the Raman scattering. This is closely related to the unique forms of $\alpha'_{ix,xx}$ in AA′B′B stacking. For even $N$, $S_{N/2}$ mode exhibits the largest intensity, while for odd $N$, $S_{(N-1)/2}$ and $S_{(N+1)/2}$ modes exhibit the largest intensities (Fig. 5f). With increasing $N$, the frequency separation of $S_{(N-1)/2}$ and $S_{(N+1)/2}$ peaks decreases, eventually approaching the $S_{N/2}$ peak in the bulk.

The intensities of S modes computed from our interlayer bond polarizability model in Fig. 5

show distinct stacking dependence at any thickness, which can serve as a guiding principle for stacking determination of MX$_2$. For example, in 3L MoSe$_2$ (Fig. 5b), only the highest-frequency mode S$_1$ is observed in AA′ (or AB) stacking type, only the lowest-frequency S$_2$ mode is observed in ABC stacking type, while both S$_1$ and S$_2$ modes can be observed with nearly equal intensities in AA′B′B stacking type; in 4L MoSe$_2$ (Fig. 5c), only the highest-frequency mode S$_1$ and the third highest-frequency mode S$_3$ are observed in AA′ (or AB) stacking type with $I(S_1) > I(S_3)$, only the lowest-frequency mode S$_3$ and the third lowest-frequency mode S$_1$ are observed in ABC stacking type with $I(S_3) > I(S_1)$, while the middle mode S$_2$ dominate in AA′B′B stacking type. Similar trends are found in other MX$_2$ such as MoS$_2$ and WSe$_2$.[49,50,52,53] Note that the crucial parameters are set as $|\beta_1| = 2.32$, $\phi_1 = 118.74°$, $|\beta_2| = 1.00$, and $\phi_2 = 30.00°$ to reproduce the experimentally observed equal intensities for S$_1$ and S$_2$ modes of 3L MoSe$_2$ in AA′B′B stacking type.[49] However, the magnitudes and phase angles of $\beta_1$ and $\beta_2$ vary among different MX$_2$ and laser wavelengths, which can give rise to different relative intensities between S$_1$ and S$_2$ modes, as observed for 3L MoS$_2$ in AA′B′B (mixed) stacking type by Lee et al.[53] Thus careful parameter fitting is needed to quantitatively account for experimental Raman measurements.

Turning to the $z$ direction in $N$-layer MX$_2$, for an interior layer $i$, the $z$ components of the two normalized interlayer bond vectors always assume a relation $\hat{R}_{iz,i-1} = -\hat{R}_{iz,i+1}$ regardless of the in-plane stacking details. Similar to graphene, for AA′ or AB or ABC stacking, $\alpha'_{1z,xx} = \gamma$, $\alpha'_{Nz,xx} = -\gamma$, while for all interior layers, $\alpha'_{iz,xx} = C^*(\hat{R}_{iz,i-1} + \hat{R}_{iz,i+1}) = 0$. Here $\gamma = \gamma_1$ for AA′ stacking, while $\gamma = \gamma_2$ for AB or ABC stacking. For AA′B′B stacking, the situation is again more complex due to the stacking mixture: $\alpha'_{1z,xx} = \gamma_1$, $\alpha'_{Nz,xx} = -\gamma_2$ for odd $N$ or $\alpha'_{Nz,xx} = -\gamma_1$ for even $N$, and $\alpha'_{2z,xx} = -\gamma_1 + \gamma_2$, $\alpha'_{3z,xx} = -\gamma_2 + \gamma_1, \ldots$, where for an interior layer $i$, $\alpha'_{iz,xx} = -\alpha'_{(i+1)z,xx}$ and thus $\alpha'_{iz,xx} = \alpha'_{(i+2)z,xx}$ (see Eq. S18 in Section S2 in SI). With the normalized layer displacements $\Delta r_{iz}$ given by Eq. 12 and the frequencies of all $N-1$ B modes given by Eq. 11, we can obtain the intensities of the B modes in all four stacking polytypes for $N$-layer MX$_2$.

Taking MoS$_2$ as an example, where the frequency of the bulk B mode is $\omega(B_{bulk}) \approx 48.1$ cm$^{-1}$, we computed the thickness dependence of B modes in various stackings as shown in Fig. 6. In AA′ or AB or ABC stacking, the forms of $\alpha'_{iz,xx}$ are the same as those found for graphene. Therefore, according to Eq. 18, the polarizability change by a breathing vibration, $\Delta\alpha_{xx}$, is only related to the displacement difference between the top and bottom layers, and the largest value occurs for B$_{N-1}$ (the lowest-frequency B branch) that has the largest opposite displacements of layer 1 and layer $N$. Consequently, for the $N-1$ B modes of $N$-layer MoS$_2$ in these three stackings, starting from the lowest-frequency one, only B$_{N-1}$, B$_{N-3}$, B$_{N-5}$, ... can be observed with an intensity trend $I(B_{N-1}) > I(B_{N-3}) > I(B_{N-5}) > ...$ (Fig. 6a). Such trend is the same as that for graphene in Fig. 3e. The theoretical results in Fig. 6a shed light on why only the lowest-frequency, third lowest-frequency, and fifth lowest-frequency B modes (B$_{N-1}$, B$_{N-3}$ and B$_{N-5}$) can be observed for natural AA′-stacked (2H-stacked) MoS$_2$ and WSe$_2$ samples at different thicknesses.[36,39] Additionally, the prediction of $I(B_{N-1}) > I(B_{N-3}) > I(B_{N-5})$ from our model is well consistent with experimental data.[36,39] Similar to the S modes, although B$_{N-3}$, B$_{N-5}$, B$_{N-7}$, ... are in principle observable starting from 4L, 6L, 8L, ..., respectively, experimentally they could be too weak to be observed at all, or could be strong enough to be observed only at thicker layers. This is particularly true for B$_{N-5}$, B$_{N-7}$, etc..[36,39] For bulk MoS$_2$ in AA′ or AB or ABC stacking, there are no exterior layers and for any layer we have $\hat{R}_{iz,i-1} = -\hat{R}_{iz,i+1}$, giving rise to $\alpha'_{iz,xx} = C^*(\hat{R}_{iz,i-1} + \hat{R}_{iz,i+1}) = 0$. Consequently none of breathing vibrations change the polarizability, and the intensities of the B modes are zero for the bulk (Fig. 6a). This can explain why the B mode is Raman inactive in 2H-stacked bulk MX$_2$.[36,39] Note that although AB (3R′) and ABC (3R) stackings show the same intensities of B modes, AA′ (2H) stacking exhibits systematically lower intensities of B modes than AB or ABC stacking, i.e., $|\gamma_1| < |\gamma_2|$. In contrast, AA′ stacking exhibits systemically higher intensities of S modes than AB stacking, i.e., $|\beta_1| > |\beta_2|$, as discussed above. Such opposite behaviors between the B and S modes were reported by both experimental measurements and first-principles calculations for MoS$_2$ and WSe$_2$.[49,53,66] Thus the intensities of B modes can also help to differentiate AA′ (2H) from AB (3R′) and ABC (3R), while AB (3R′) and ABC (3R) require the S modes for differentiation as discussed before.

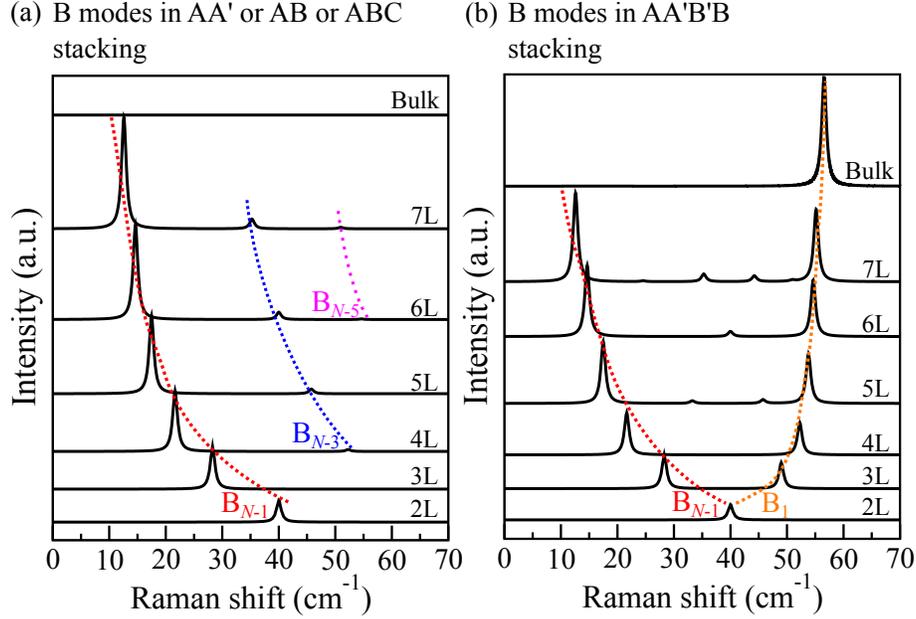

FIG. 6. (Color online) Calculated thickness dependence of B modes of MoS$_2$ in (a) AA′ (2H) or AB (3R′) or ABC (3R) stacking, and (b) AA′B′B (2H-3R mixed) stacking. Similar results apply to other group-6 MX$_2$. Dashed lines indicate the frequency evolution trends of the modes. In (a), the B modes in AA′ stacking exhibit systemically lower intensities than those in AB or ABC stacking, though the trends are the same. For 2L in (b), the lowest-frequency B branch B$_{N-1}$ is equivalent to the highest-frequency B branch B$_1$.

In AA′B′B (2H-3R mixed) stacking, for the interior layers, since $\alpha'_{iz,xx}$ is no longer zero and every adjacent layers have opposite $\alpha'_{iz,xx}$, the highest-frequency branch B$_1$ that involves opposite vibrations between every adjacent layers has non-zero intensity now, besides the lowest-frequency branch B$_{N-1}$ (Fig. 6b). They are two major peaks among the B modes. As a general rule from linear chain model, for the highest-frequency branch B$_1$, its frequency increases with increasing thickness and approaches the bulk B mode, while for the lowest-frequency branch B$_{N-1}$, its frequency decreases with increasing thickness and reaches zero in the bulk (thus cannot be observed). Interestingly, unlike AA′ or AB or ABC stacking where the bulk B mode has zero intensity as $\alpha'_{iz,xx} = 0$ for any layer, the stacking mixture in AA′B′B gives rise to a distinct B peak in the bulk (Fig. 6), as the forms of $\alpha'_{iz,xx}$ in the bulk assume an repeated pattern of $-\gamma_2 + \gamma_1$ and $-\gamma_1 + \gamma_2$. Obviously, the B modes can also be fingerprints to identify the mixed stacking AA′B′B. Note that if we have $\gamma_1 = \gamma_2$ (i.e., uniform interlayer bond polarizabilities between every layers), Fig. 6b would recover to Fig. 6a, and the B$_1$ branch would disappear.

### C. Application of the model to other 2D materials

Besides graphene and MoS$_2$, our interlayer bond polarizability model can also be applied to many other 2D materials. In NbSe$_2$,[46] an 2D superconductor with a natural stacking type of AB′ (*eclipsed* with Nb over Nb, while Se atoms over the hexagonal centers, see Fig. S1), its interlayer bond vectors are similar to AA′ or AB stacking types in Fig. 4a, since the stacking sequence also repeats after two layers. For an interior layer $i$, we have $\hat{R}_{ix,i-1} = \hat{R}_{ix,i+1}$ in the $x$ direction, but $\hat{R}_{iz,i-1} = -\hat{R}_{iz,i+1}$ in the $z$ direction. Therefore, the polarizability derivative for each layer's displacement along the $x$ direction assumes $\alpha'_{1x,xx} = \beta$, $\alpha'_{2x,xx} = -2\beta$, $\alpha'_{3x,xx} = 2\beta$, $\alpha'_{4x,xx} = -2\beta$, $\alpha'_{5x,xx} = 2\beta$, ..., $\alpha'_{Nx,xx} = \beta$ (odd $N$) or $\alpha'_{Nx,xx} = -\beta$ (even $N$); while the polarizability derivative for each layer's displacement along the $z$ direction assumes $\alpha'_{1z,xx} = \gamma$, $\alpha'_{Nz,xx} = -\gamma$, and $\alpha'_{iz,xx} = 0$ for all interior layers. Following the similar procedures employed before, the polarizability change in $N$-layer NbSe$_2$ samples due to the shear or breathing vibrations is

$$\Delta\alpha_{xx}(\text{AB}',\text{S}) = \beta(\Delta r_{Nx} - \Delta r_{1x}) + 2\beta \sum_{i=1,3,5}^{m} (\Delta r_{ix} - \Delta r_{(i+1)x});$$

$$\Delta\alpha_{xx}(\text{AB}',\text{B}) = \gamma(\Delta r_{1z} - \Delta r_{Nz}), \tag{23}$$

respectively. These equations are the same as the formula for AB stacking in Eq. 17 and Eq. 18. Similar to AA$'$ or AB stacking, for AB$'$ stacking, the highest-frequency S branch (S$_1$) exhibits the largest intensity among all S modes, and its frequency increases with increasing thickness and eventually reaches the bulk Raman-active S mode; in contrast, the lowest-frequency B branch (B$_{N-1}$) dominates among the B modes, and its frequency decreases with increasing thickness and reaches zero in the bulk (Fig. 7a). As discussed before, the bulk B mode has zero intensity. The theoretical results from our model are in agreement with experimental data on NbSe$_2$ in Ref. 46. Obviously, the evolution of LF Raman spectra with thickness in AB$'$-stacked NbSe$_2$ in Fig. 7a is very similar to that of AB-stacked graphene (Figs. 3c and 3e) and that of AA$'$- or AB- stacked MoS$_2$ or MoSe$_2$ (Figs. 5d and 6a). This suggests that for AA$'$, AB, and AB$'$ stacking types, the trends of LF Raman response to stacking and thickness are the same as Fig. 7a, independent of the 2D materials and their specific structures and symmetries, since each layer can be simplified as a single object and the three stackings share the similar interlayer bond vectors, according to our model. Our analysis is further validated by recent Raman measurements on h-BN in natural AA$'$ stacking (*eclipsed* with B over N),[77] where the highest-frequency S branch (S$_1$) was observed as it exhibits the largest intensity among the S modes, similar to the S$_1$ peak in Fig. 7a.

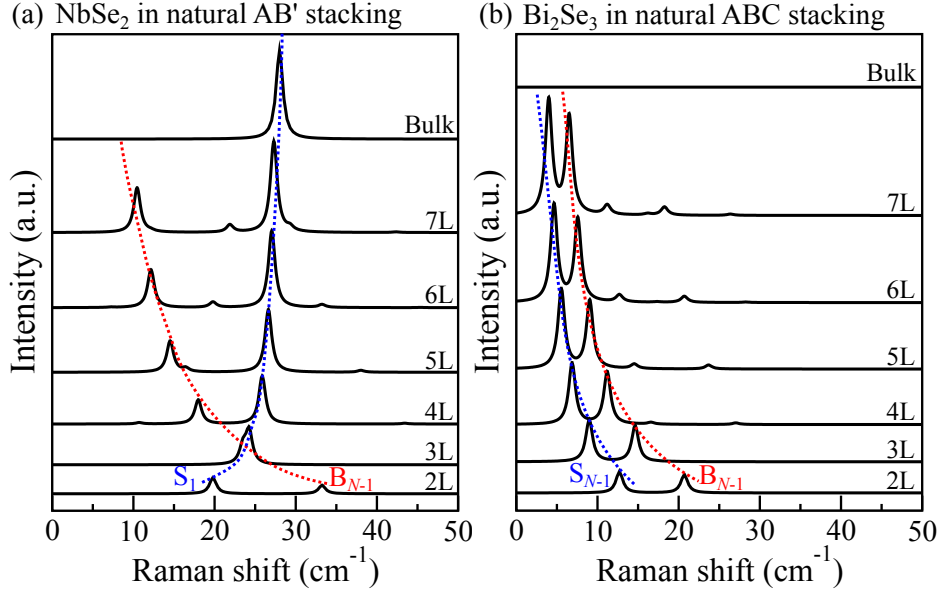

FIG. 7. (Color online) Calculated LF Raman scattering of (a) NbSe$_2$ in natural AB$'$ stacking and (b) Bi$_2$Se$_3$ in natural ABC stacking at different thicknesses from bilayer to bulk, based on the interlayer bond polarizability model. Dashed lines indicate the frequency evolution trends of the major S and B peaks. For 3L NbSe$_2$ in (a), the S and B peaks almost overlap with each other. As a general rule from linear chain model, S$_{N-1}$ (B$_{N-1}$) is the lowest-frequency S (B) branch, and the frequency decreases with increasing thickness; in contrast, S$_1$ (B$_1$) is the highest-frequency S (B) branch, and the frequency increases with increasing thickness. The theoretical results in (a) and (b) explain experimental data from Ref. 46 and Ref. 44, respectively.

In Bi$_2$Se$_3$,[44] a 2D topological insulator with a natural stacking type of ABC, its interlayer bond vectors are similar to the ABC stacking type in Fig. 4b. For an interior layer $i$, we have $\hat{R}_{ix,i-1} = -\hat{R}_{ix,i+1}$ in the $x$ direction, and similarly $\hat{R}_{iz,i-1} = -\hat{R}_{iz,i+1}$ in the $z$ direction. Thus, $\alpha'_{ix,xx}$ and $\alpha'_{iz,xx}$ assume similar forms: $\alpha'_{1x,xx} = \beta$, $\alpha'_{Nx,xx} = -\beta$, and $\alpha'_{ix,xx} = 0$ for all interior layers; $\alpha'_{1z,xx} = \gamma$, $\alpha'_{Nz,xx} = -\gamma$, and $\alpha'_{iz,xx} = 0$ for all interior layers. Following the similar procedures aforementioned, the polarizability change of $N$-layer Bi$_2$Se$_3$ by the shear or breathing vibrations is similar:

$$\Delta\alpha_{xx}(\text{ABC}, \text{S}) = \beta(\Delta r_{1x} - \Delta r_{Nx});$$
$$\Delta\alpha_{xx}(\text{ABC}, \text{B}) = \gamma(\Delta r_{1z} - \Delta r_{Nz}). \quad (24)$$

These equations are the same as the formula obtained for ABC stacking in Eq. 17 and Eq. 18. It is apparent that the S and B modes exhibit the same behaviors. Similar to ABC-stacked graphene (Figs. 3d and 3e), and ABC-stacked MoS$_2$ and MoSe$_2$ (Figs. 5e and 6a), for ABC-stacked Bi$_2$Se$_3$, the lowest-frequency S branch (S$_{N-1}$) and B branch (B$_{N-1}$) dominate among the S and B modes, respectively. Their frequencies decrease with increasing thickness and disappear in the bulk (Fig. 7b). For ABC stacking, both the bulk S and B modes cannot be observed as discussed previously. The theoretical results from our model in Fig. 7b are consistent with experimental data in Ref. 44, and similar results can be found for ABC-stacked Bi$_2$Te$_3$ as well. Obviously, the evolution of LF Raman spectra with thickness in ABC stacking is also generalized, independent of the specific 2D materials.

## IV. CONCLUSIONS

In conclusion, a simplified and generalized interlayer bond polarizability model has been developed to understand and predict LF Raman spectra in any thickness and stacking for diverse 2D materials. Additionally, a general strategy is also proposed to unify the stacking nomenclature for 2D materials. Our model successfully explains a wide range of existing experimental data for graphene, $MoS_2$, $MoSe_2$, $WSe_2$, $NbSe_2$, $Bi_2Se_3$, h-BN. It is also expected to be applicable to many other 2D materials. The key for the simplicity and generalization of our model is that each layer is treated as a single object with no need of intralayer structural details, only the interlayer bond vectors and polarizabilities are required to determine Raman intensities of both shear and breathing modes. This allows both experimentalists and theorists to quickly diagnose their data without time-consuming first-principles Raman calculations. This is particularly appealing for thick samples with complex stacking sequences. Our work reveals the fundamental mechanism of stacking-dependent LF Raman response, which is that different stacking types can change the interlayer bond vectors and/or bond polarizabilities. The LF Raman modes can be effective stacking fingerprints and some general rules are summarized below as guidelines.

(a) In AA′, AB, and AB′ stacking types for which the stacking sequence repeats after two layers, they share similar interlayer bond vectors. For an interior layer $i$, we have $\hat{R}_{ix,i-1} = \hat{R}_{ix,i+1}$ in the $x$ direction. Starting from the highest-frequency S mode, only $S_1$, $S_3$, $S_5$, … can be observed with an intensity trend $I(S_1) > I(S_3) > I(S_5) > \ldots$, and their frequencies increase with increasing thickness. Although the S modes exhibit similar trends among these stacking types, the intensities can be quite different since the interlayer bond polarizabilities are stacking dependent, which may help for their differentiation.

(b) In ABC stacking type for which the stacking sequence repeats after three layers, the $x$ (in-plane) components of the interlayer bond vectors are changed after the in-plane stacking variation. For an interior layer $i$, we have $\hat{R}_{ix,i-1} = -\hat{R}_{ix,i+1}$ in the $x$ direction. Subsequently, the trend is the opposite and starting from the lowest-frequency S mode, only $S_{N-1}$, $S_{N-3}$, $S_{N-5}$, … can be observed with an intensity trend $I(S_{N-1}) > I(S_{N-3}) > I(S_{N-5}) > \ldots$, and their frequencies decrease with increasing thickness. It follows that the S modes can be fingerprints to distinguish ABC from AA′, AB, and AB′ stacking types.

(c) However, the *z* (out-of-plane) components of the interlayer bond vectors do not change with the in-plane stacking variation, and for an interior layer *i* we always have $\hat{R}_{iz,i-1} = -\hat{R}_{iz,i+1}$ in the *z* direction, regardless of the stacking types. This relation is similar to that in ABC stacking for the *x* direction. Consequently, for all AA′, AB, AB′, and ABC stacking types, starting from the lowest-frequency one, only $B_{N-1}$, $B_{N-3}$, $B_{N-5}$, ... can be observed with an intensity trend $I(B_{N-1}) > I(B_{N-3}) > I(B_{N-5}) > ...$, and their frequencies decrease with increasing thickness. Such trend is similar to that of S modes in ABC stacking above. Although the B modes exhibit similar trends among these stacking types, the intensities can be also quite different, which may be used for their differentiation. For instance, from AA′ to AB (i.e., 2H to 3R) stacking in bilayer $MoS_2$ and $WSe_2$, the S mode intensity is reduced while the B mode intensity is enhanced. Such opposite behaviors can be signatures for stacking determination.

(d) Finally, besides AA′, AB, AB′, and ABC stacking types, stacking mixtures, which could occur during sample growth and preparation, give rise to non-uniform interlayer bond vectors and polarizabilities. For example, in AA′B′B stacking studied in this work, the mixture of AA′ and AB stackings lead to more complex LF Raman behaviors. Instead of the highest-frequency $S_1$ or lowest-frequency $S_{N-1}$, the S modes in the middle dominate among the S modes. Additionally, instead of only the lowest-frequency $B_{N-1}$ being dominant among the B modes, the highest-frequency branch $B_1$ is also distinctively noticeable. Therefore, the LF modes can also facilitate the identification of stacking mixtures.

**ACKNOWLEDGMENTS**

L.L. was supported by Eugene P. Wigner Fellowship at the Oak Ridge National Laboratory. L.L. A.A.P and B.G.S. acknowledge work at the Center for Nanophase Materials Sciences, which is a DOE Office of Science User Facility. V.M. acknowledges the support by NSF and the Office of Naval Research.

* [liangl1@ornl.gov](mailto:liangl1@ornl.gov)

† [meuniv@rpi.edu](mailto:meuniv@rpi.edu)

# Supporting Information

# Interlayer bond polarizability model for stacking-dependent low-frequency Raman scattering in layered materials

Liangbo Liang,[1,2,*] Alexander A. Puretzky,[1] Bobby G. Sumpter,[1] and Vincent Meunier[2,†]

[1]*Center for Nanophase Materials Sciences,*
*Oak Ridge National Laboratory, Oak Ridge, Tennessee 37831, USA*
[2]*Department of Physics, Applied Physics, and Astronomy,*
*Rensselaer Polytechnic Institute, Troy, New York 12180, USA*

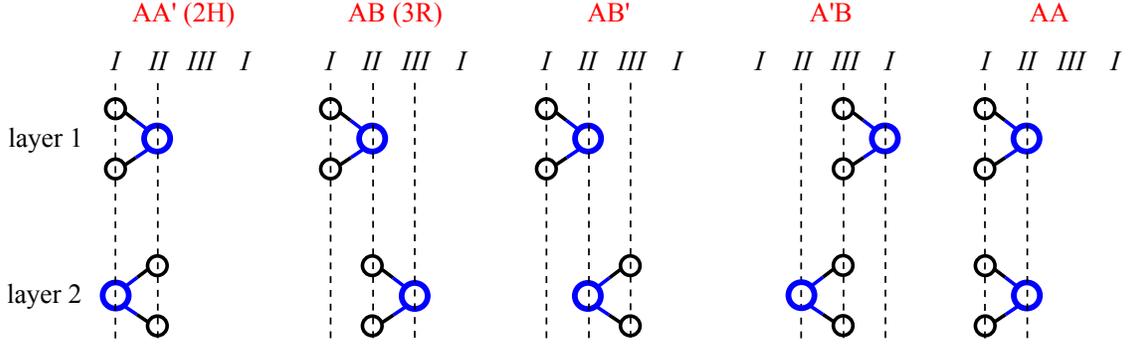

FIG. S1. Side views of the five possible high-symmetry stacking patterns in bilayer TMDs $MX_2$ like $MoS_2$. The blue (black) circles represent metal M (chalcogenide X) atoms. In the honeycomb lattice, there are three atomic coordinates: *I*: 0, 0; *II*: 1/3, 2/3; *III*: 2/3, 1/3. In bilayer $MX_2$, AA′ (corresponding to bulk 2H stacking) and AB (corresponding to bulk 3R stacking) are stable and can be commonly found in natural and synthetic samples.

## S1. GENERALIZED INTERLAYER BOND POLARIZABILITY MODEL

In this section, we present the detailed derivation process of the generalized bond polarizability model.[1–4] The Raman intensity of a phonon mode $k$ is given by[1,4,5]

$$I(k) \propto \left| e_i \cdot \tilde{R}(k) \cdot e_s^T \right|^2 \propto \left| \sum_{\mu\nu} e_{i,\mu} e_{s,\nu} \Delta\alpha_{\mu\nu}(k) \right|^2, \quad (S1)$$

where $\tilde{R}(k)$ is the (3×3) Raman tensor of the phonon mode $k$, subscripts $\mu$ and $\nu$ indicate Cartesian components ($x$, $y$ or $z$) of the tensor, and $e_i$ and $e_s$ are the unit vectors for the polarization of the incident and scattered light, respectively. The Raman tensor element

$$\Delta\alpha_{\mu\nu}(k) = \sum_{j\gamma} \left[ \frac{\partial \alpha_{\mu\nu}}{\partial r_{j\gamma}} \right]_0 \Delta r_{j\gamma}(k), \quad (S2)$$

where $r_{j\gamma}$ is the position of atom $j$ along direction $\gamma$ ($x$, $y$ or $z$) in equilibrium, $\left[ \frac{\partial \alpha_{\mu\nu}}{\partial r_{j\gamma}} \right]_0$ is the derivative of the electronic polarizability tensor element $\alpha_{\mu\nu}$ with respect to the atomic displacement from the equilibrium configuration, and $\Delta r_{j\gamma}(k)$ is the eigen-displacement of atom $j$ along direction $\gamma$ in the phonon mode $k$ (i.e., the eigenvector of the mass-normalized dynamic matrix).[5] One can see that the Raman tensor of the phonon mode $k$ is proportional to the change of the polarizability by its vibration. According to the empirical bond polarizability model, the polarizability of the system can be approximated by a sum of individual bond polarizabilities from different bonds:[1,4]

$$\alpha_{\mu\nu} = \frac{1}{2}\sum_{iB}\left[\frac{\alpha_{\parallel,B}+2\alpha_{\perp,B}}{3}\delta_{\mu\nu} + (\alpha_{\parallel,B}-\alpha_{\perp,B})\left(\frac{R_{i\mu,B}R_{i\nu,B}}{R_{i,B}^2} - \frac{1}{3}\delta_{\mu\nu}\right)\right], \quad (S3)$$

where $B$ indicates a bond connected to atom $i$, the boldface $\boldsymbol{R}_{i,B}$ is the corresponding bond vector connecting atom $i$ to one of its neighbor atoms $i'$, $R_{i\mu,B}$ is the $\mu$ ($x$, $y$ or $z$) component of $\boldsymbol{R}_{i,B}$, and $R_{i,B}$ is the length of $\boldsymbol{R}_{i,B}$. $\alpha_{\parallel,B}$ and $\alpha_{\perp,B}$ are the bond polarizabilities for the bond $B$ in the directions parallel and perpendicular to the bond, respectively. They depend on the bond $R_{i,B}$ and therefore,

$$\begin{aligned}\frac{\partial \alpha_{\parallel,B}}{\partial r_{j\gamma}} &= \frac{\partial \alpha_{\parallel,B}}{\partial R_{i,B}}\frac{\partial R_{i,B}}{\partial r_{j\gamma}} = \alpha'_{\parallel,B}\frac{\partial R_{i,B}}{\partial r_{j\gamma}}\\ \frac{\partial \alpha_{\perp,B}}{\partial r_{j\gamma}} &= \frac{\partial \alpha_{\perp,B}}{\partial R_{i,B}}\frac{\partial R_{i,B}}{\partial r_{j\gamma}} = \alpha'_{\perp,B}\frac{\partial R_{i,B}}{\partial r_{j\gamma}},\end{aligned} \quad (S4)$$

where $\alpha'_{\parallel,B}$ and $\alpha'_{\perp,B}$ are the radial derivatives of the bond polarizabilities with respect to the bond length. The values of $\alpha_{\parallel,B}$, $\alpha_{\perp,B}$, $\alpha'_{\parallel,B}$ and $\alpha'_{\perp,B}$ are functions of the bond length, and usually determined by fitting with experimental Raman intensities.[1–4]

To obtain $\frac{\partial R_{i,B}}{\partial r_{j\gamma}}$, we need to use the following relations: $\boldsymbol{R}_{i,B} = \vec{r}_{i'} - \vec{r}_i, \boldsymbol{R}_{i',B} = \vec{r}_i - \vec{r}_{i'} = -\boldsymbol{R}_{i,B}, R_{i,B} = \sqrt{\sum_\eta (r_{i'\eta} - r_{i\eta})^2} = R_{i',B}$. Subsequently,

$$\sum_{j\gamma} \frac{\partial R_{i,B}}{\partial r_{j\gamma}} \Delta r_{j\gamma}(k) = \sum_{j\gamma} \frac{\partial \sqrt{\sum_\eta (r_{i'\eta} - r_{i\eta})^2}}{\partial r_{j\gamma}} \Delta r_{j\gamma}(k) = \sum_{j\gamma} \frac{1}{2} \frac{1}{\sqrt{\sum_\eta (r_{i'\eta} - r_{i\eta})^2}} \frac{\partial \sum_\eta (r_{i'\eta} - r_{i\eta})^2}{\partial r_{j\gamma}} \Delta r_{j\gamma}(k)$$

$$= \sum_{j\gamma} \frac{1}{2} \frac{1}{\sqrt{\sum_\eta (r_{i'\eta} - r_{i\eta})^2}} \left( \sum_\eta 2(r_{i'\eta} - r_{i\eta}) \frac{\partial (r_{i'\eta} - r_{i\eta})}{\partial r_{j\gamma}} \right) \Delta r_{j\gamma}(k) \qquad (\eta \text{ or } \gamma = x, y, z)$$

$$= \sum_{j\gamma} \frac{1}{2} \frac{1}{\sqrt{\sum_\eta (r_{i'\eta} - r_{i\eta})^2}} \left( \sum_\eta 2(r_{i'\eta} - r_{i\eta})(\delta_{i'j} \delta_{\eta\gamma} - \delta_{ij} \delta_{\eta\gamma}) \right) \Delta r_{j\gamma}(k)$$

$$= \sum_{j\gamma} \frac{1}{R_{i,B}} \left[ (r_{i'\gamma} - r_{i\gamma})(\delta_{i'j} - \delta_{ij}) \right] \Delta r_{j\gamma}(k) = \sum_{j\gamma} \frac{(r_{i'\gamma} - r_{i\gamma})\delta_{i'j}}{R_{i,B}} \Delta r_{j\gamma}(k) - \sum_{j\gamma} \frac{(r_{i'\gamma} - r_{i\gamma})\delta_{ij}}{R_{i,B}} \Delta r_{j\gamma}(k)$$

$$= \sum_\gamma \frac{(r_{i'\gamma} - r_{i\gamma})}{R_{i,B}} \Delta r_{i'\gamma}(k) - \sum_\gamma \frac{(r_{i'\gamma} - r_{i\gamma})}{R_{i,B}} \Delta r_{i\gamma}(k) = -\sum_\gamma \frac{(r_{i\gamma} - r_{i'\gamma})}{R_{i,B}} \Delta r_{i'\gamma}(k) - \sum_\gamma \frac{(r_{i'\gamma} - r_{i\gamma})}{R_{i,B}} \Delta r_{i\gamma}(k)$$

$$= -\frac{(\vec{r}_i - \vec{r}_{i'})}{R_{i,B}} \cdot \Delta \vec{r}_{i'}(k) - \frac{(\vec{r}_{i'} - \vec{r}_i)}{R_{i,B}} \cdot \Delta \vec{r}_i(k) = -\frac{\boldsymbol{R}_{i',B}}{R_{i',B}} \cdot \Delta \vec{r}_{i'}(k) - \frac{\boldsymbol{R}_{i,B}}{R_{i,B}} \cdot \Delta \vec{r}_i(k) \qquad (R_{i',B} = R_{i,B})$$

$$\Downarrow$$

$$\sum_{iB} \sum_{j\gamma} \frac{\partial R_{i,B}}{\partial r_{j\gamma}} \Delta r_{j\gamma}(k) = -\sum_{iB} \left( \frac{\boldsymbol{R}_{i',B}}{R_{i',B}} \cdot \Delta \vec{r}_{i'}(k) \right) - \sum_{iB} \left( \frac{\boldsymbol{R}_{i,B}}{R_{i,B}} \cdot \Delta \vec{r}_i(k) \right) = -2 \sum_{iB} \left( \frac{\boldsymbol{R}_{i,B}}{R_{i,B}} \cdot \Delta \vec{r}_i(k) \right),$$

(S5)

where $\sum_{iB} \left( \frac{\boldsymbol{R}_{i',B}}{R_{i',B}} \cdot \Delta \vec{r}_{i'}(k) \right) = \sum_{ii'B} \left( \frac{\boldsymbol{R}_{i',B}}{R_{i',B}} \cdot \Delta \vec{r}_{i'}(k) \right) = \sum_{i'B} \left( \frac{\boldsymbol{R}_{i',B}}{R_{i',B}} \cdot \Delta \vec{r}_{i'}(k) \right) = \sum_{iB} \left( \frac{\boldsymbol{R}_{i,B}}{R_{i,B}} \cdot \Delta \vec{r}_i(k) \right)$.
Furthermore,

$$\sum_{iB} \sum_{j\gamma} \frac{\partial}{\partial r_{j\gamma}} \left( \frac{1}{R_{i,B}^2} \right) \Delta r_{j\gamma}(k) = \sum_{iB} \sum_{j\gamma} \left( -\frac{2}{R_{i,B}^3} \frac{\partial R_{i,B}}{\partial r_{j\gamma}} \Delta r_{j\gamma}(k) \right)$$

$$= \sum_{iB} -\frac{2}{R_{i,B}^3} \left( \sum_{j\gamma} \frac{\partial R_{i,B}}{\partial r_{j\gamma}} \Delta r_{j\gamma}(k) \right)$$

$$= \sum_{iB} -\frac{2}{R_{i,B}^3} \left( -\frac{\boldsymbol{R}_{i',B}}{R_{i',B}} \cdot \Delta \vec{r}_{i'}(k) - \frac{\boldsymbol{R}_{i,B}}{R_{i,B}} \cdot \Delta \vec{r}_i(k) \right) \text{ (see Eq. S5)}$$

$$= 2 \sum_{iB} \left( \frac{\boldsymbol{R}_{i',B}}{R_{i',B}^4} \cdot \Delta \vec{r}_{i'}(k) + \frac{\boldsymbol{R}_{i,B}}{R_{i,B}^4} \cdot \Delta \vec{r}_i(k) \right) \qquad (R_{i',B} = R_{i,B})$$

$$= 4 \sum_{iB} \frac{\boldsymbol{R}_{i,B}}{R_{i,B}^4} \cdot \Delta \vec{r}_i(k),$$

(S6)

where similarly $\sum_{iB} \frac{\boldsymbol{R}_{i',B}}{R_{i',B}^4} \cdot \Delta \vec{r}_{i'}(k) = \sum_{iB} \frac{\boldsymbol{R}_{i,B}}{R_{i,B}^4} \cdot \Delta \vec{r}_i(k)$. In addition, the $\mu$ ($x$, $y$ or $z$) component of $\boldsymbol{R}_{i,B}$ is $R_{i\mu,B} = r_{i'\mu} - r_{i\mu}$, and similarly $R_{i\nu,B} = r_{i'\nu} - r_{i\nu}$. It follows that

$$\sum_{j\gamma} \frac{\partial (R_{i\mu,B} R_{i\nu,B})}{\partial r_{j\gamma}} \Delta r_{j\gamma}(k) = \sum_{j\gamma} \frac{\partial (r_{i'\mu} - r_{i\mu})}{\partial r_{j\gamma}} R_{i\nu,B} \Delta r_{j\gamma}(k) + \sum_{j\gamma} R_{i\mu,B} \frac{\partial (r_{i'\nu} - r_{i\nu})}{\partial r_{j\gamma}} \Delta r_{j\gamma}(k)$$

$$= \sum_{j\gamma} (\delta_{i'j} \delta_{\mu\gamma} - \delta_{ij} \delta_{\mu\gamma}) R_{i\nu,B} \Delta r_{j\gamma}(k) + \sum_{j\gamma} R_{i\mu,B} (\delta_{i'j} \delta_{\nu\gamma} - \delta_{ij} \delta_{\nu\gamma}) \Delta r_{j\gamma}(k)$$

$$= \sum_j (\delta_{i'j} - \delta_{ij}) R_{i\nu,B} \Delta r_{j\mu}(k) + \sum_j R_{i\mu,B} (\delta_{i'j} - \delta_{ij}) \Delta r_{j\nu}(k)$$

$$= \left( R_{i\nu,B} \Delta r_{i'\mu}(k) - R_{i\nu,B} \Delta r_{i\mu}(k) \right) + \left( R_{i\mu,B} \Delta r_{i'\nu}(k) - R_{i\mu,B} \Delta r_{i\nu}(k) \right)$$

$$= \left( -R_{i'\nu,B} \Delta r_{i'\mu}(k) - R_{i\nu,B} \Delta r_{i\mu}(k) \right) + \left( -R_{i'\mu,B} \Delta r_{i'\nu}(k) - R_{i\mu,B} \Delta r_{i\nu}(k) \right) \qquad (R_{i\nu,B} = -R_{i'\nu,B}; R_{i\mu,B} = -R_{i'\mu,B})$$

$$\Downarrow$$

$$\sum_{iB} \sum_{j\gamma} \frac{\partial (R_{i\mu,B} R_{i\nu,B})}{\partial r_{j\gamma}} \Delta r_{j\gamma}(k) = \sum_{iB} \left( -R_{i'\nu,B} \Delta r_{i'\mu}(k) - R_{i\nu,B} \Delta r_{i\mu}(k) \right) + \sum_{iB} \left( -R_{i'\mu,B} \Delta r_{i'\nu}(k) - R_{i\mu,B} \Delta r_{i\nu}(k) \right)$$

$$= -2 \sum_{iB} \left( R_{i\nu,B} \Delta r_{i\mu}(k) + R_{i\mu,B} \Delta r_{i\nu}(k) \right), \tag{S7}$$

where similarly, $\sum_{iB} R_{i'\nu,B} \Delta r_{i'\mu}(k) = \sum_{ii'B} R_{i'\nu,B} \Delta r_{i'\mu}(k) = \sum_{i'B} R_{i'\nu,B} \Delta r_{i'\mu}(k) = \sum_{iB} R_{i\nu,B} \Delta r_{i\mu}(k)$, and $\sum_{iB} R_{i'\mu,B} \Delta r_{i'\nu}(k) = \sum_{iB} R_{i\mu,B} \Delta r_{i\nu}(k)$.

With Eqs. S4, S5, S6 and S7, we then substitute Eq. S3 into Eq. S2, which yields the Raman tensor element

$$\Delta\alpha_{\mu\nu}(k) = \sum_{j\gamma}\frac{\partial}{\partial r_{j\gamma}}\left\{\frac{1}{2}\sum_{iB}\left[\frac{\alpha_{\parallel,B}+2\alpha_{\perp,B}}{3}\delta_{\mu\nu} + (\alpha_{\parallel,B}-\alpha_{\perp,B})\left(\frac{R_{i\mu,B}R_{i\nu,B}}{R_{i,B}^2} - \frac{1}{3}\delta_{\mu\nu}\right)\right]\right\}\Delta r_{j\gamma}(k)$$

$$=\frac{1}{2}\sum_{iB}\left\{\left[\frac{\alpha'_{\parallel,B}+2\alpha'_{\perp,B}}{3}\delta_{\mu\nu} + (\alpha'_{\parallel,B}-\alpha'_{\perp,B})\left(\frac{R_{i\mu,B}R_{i\nu,B}}{R_{i,B}^2}-\frac{1}{3}\delta_{\mu\nu}\right)\right]\sum_{j\gamma}\frac{\partial R_{i,B}}{\partial r_{j\gamma}}\Delta r_{j\gamma}(k)\right\}$$

$$+\frac{1}{2}\sum_{iB}\left\{(\alpha_{\parallel,B}-\alpha_{\perp,B})\sum_{j\gamma}\frac{\partial}{\partial r_{j\gamma}}\left(\frac{R_{i\mu,B}R_{i\nu,B}}{R_{i,B}^2}\right)\Delta r_{j\gamma}(k)\right\}$$

$$=\frac{1}{2}\sum_{iB}\left\{\left[\frac{\alpha'_{\parallel,B}+2\alpha'_{\perp,B}}{3}\delta_{\mu\nu} + (\alpha'_{\parallel,B}-\alpha'_{\perp,B})\left(\frac{R_{i\mu,B}R_{i\nu,B}}{R_{i,B}^2}-\frac{1}{3}\delta_{\mu\nu}\right)\right]\sum_{j\gamma}\frac{\partial R_{i,B}}{\partial r_{j\gamma}}\Delta r_{j\gamma}(k)\right\}$$

$$+\frac{1}{2}\sum_{iB}\left\{(\alpha_{\parallel,B}-\alpha_{\perp,B})\left[\frac{1}{R_{i,B}^2}\sum_{j\gamma}\frac{\partial(R_{i\mu,B}R_{i\nu,B})}{\partial r_{j\gamma}}\Delta r_{j\gamma}(k) + R_{i\mu,B}R_{i\nu,B}\sum_{j\gamma}\frac{\partial}{\partial r_{j\gamma}}\left(\frac{1}{R_{i,B}^2}\right)\Delta r_{j\gamma}(k)\right]\right\}$$

$$=\frac{1}{2}\sum_{iB}\left\{\left[\frac{\alpha'_{\parallel,B}+2\alpha'_{\perp,B}}{3}\delta_{\mu\nu} + (\alpha'_{\parallel,B}-\alpha'_{\perp,B})\left(\frac{R_{i\mu,B}R_{i\nu,B}}{R_{i,B}^2}-\frac{1}{3}\delta_{\mu\nu}\right)\right]\left(-2\frac{\boldsymbol{R}_{i,B}}{R_{i,B}}\cdot\Delta\vec{r}_i(k)\right)\right\}$$

$$+\frac{1}{2}\sum_{iB}\left\{(\alpha_{\parallel,B}-\alpha_{\perp,B})\left[-\frac{2}{R_{i,B}^2}\left(R_{i\nu,B}\Delta r_{i\mu}(k)+R_{i\mu,B}\Delta r_{i\nu}(k)\right)+4R_{i\mu,B}R_{i\nu,B}\frac{\boldsymbol{R}_{i,B}}{R_{i,B}^4}\cdot\Delta\vec{r}_i(k)\right]\right\} \text{ (see Eqs. S5-S7)}$$

$$=-\sum_{iB}\left\{\frac{\boldsymbol{R}_{i,B}}{R_{i,B}}\cdot\Delta\vec{r}_i(k)\left[\frac{\alpha'_{\parallel,B}+2\alpha'_{\perp,B}}{3}\delta_{\mu\nu}+\left(\alpha'_{\parallel,B}-\alpha'_{\perp,B}\right)\left(\frac{R_{i\mu,B}R_{i\nu,B}}{R_{i,B}^2}-\frac{1}{3}\delta_{\mu\nu}\right)\right]\right\}$$

$$-\sum_{iB}\left\{\frac{\alpha_{\parallel,B}-\alpha_{\perp,B}}{R_{i,B}}\left[\frac{R_{i\nu,B}\Delta r_{i\mu}(k)+R_{i\mu,B}\Delta r_{i\nu}(k)}{R_{i,B}} - 2\frac{R_{i\mu,B}R_{i\nu,B}}{R_{i,B}^2}\frac{\boldsymbol{R}_{i,B}}{R_{i,B}}\cdot\Delta\vec{r}_i(k)\right]\right\}$$

$$=-\sum_{iB}\left\{\hat{\boldsymbol{R}}_{i,B}\cdot\Delta\vec{r}_i(k)\left[\frac{\alpha'_{\parallel,B}+2\alpha'_{\perp,B}}{3}\delta_{\mu\nu}+\left(\alpha'_{\parallel,B}-\alpha'_{\perp,B}\right)\left(\hat{R}_{i\mu,B}\hat{R}_{i\nu,B}-\frac{1}{3}\delta_{\mu\nu}\right)\right]\right\}$$

$$-\sum_{iB}\left\{\frac{\alpha_{\parallel,B}-\alpha_{\perp,B}}{R_{i,B}}\left[\hat{R}_{i\mu,B}\Delta r_{i\nu}(k)+\hat{R}_{i\nu,B}\Delta r_{i\mu}(k)-2\hat{R}_{i\mu,B}\hat{R}_{i\nu,B}\left(\hat{\boldsymbol{R}}_{i,B}\cdot\Delta\vec{r}_i(k)\right)\right]\right\}, \quad \text{(S8)}$$

where $\hat{\boldsymbol{R}}_{i,B} = \dfrac{\boldsymbol{R}_{i,B}}{R_{i,B}}$ is the equilibrium-configuration bond vector normalized to unity, $\hat{R}_{i\mu,B}$ is the $\mu$ ($x$, $y$ or $z$) component of the normalized bond vector, and $R_{i,B}$ is the bond length in equilibrium.

For an interlayer shear mode vibrating along the $x$ direction, only the $x$ component of $\Delta\vec{r}_i(k)$

can be non-zero, which yields

$$\Delta\alpha_{\mu\nu} = -\sum_{iB}\left\{\hat{R}_{ix,B}\Delta r_{ix}\left[\frac{\alpha'_{\|,B}+2\alpha'_{\perp,B}}{3}\delta_{\mu\nu} + \left(\alpha'_{\|,B}-\alpha'_{\perp,B}\right)\left(\hat{R}_{i\mu,B}\hat{R}_{i\nu,B}-\frac{1}{3}\delta_{\mu\nu}\right)\right]\right\}$$

$$-\sum_{iB}\left\{\frac{\alpha_{\|,B}-\alpha_{\perp,B}}{R_{i,B}}\left[\hat{R}_{i\mu,B}\Delta r_{ix}\delta_{\nu x}+\hat{R}_{i\nu,B}\Delta r_{ix}\delta_{\mu x}-2\hat{R}_{i\mu,B}\hat{R}_{i\nu,B}\left(\hat{R}_{ix,B}\Delta r_{ix}\right)\right]\right\}$$

$$=-\sum_{iB}\left\{\hat{R}_{ix,B}\left[\frac{\alpha'_{\|,B}+2\alpha'_{\perp,B}}{3}\delta_{\mu\nu}+\left(\alpha'_{\|,B}-\alpha'_{\perp,B}\right)\left(\hat{R}_{i\mu,B}\hat{R}_{i\nu,B}-\frac{1}{3}\delta_{\mu\nu}\right)\right]\right.$$

$$\left.+\frac{\alpha_{\|,B}-\alpha_{\perp,B}}{R_{i,B}}\left[\hat{R}_{i\mu,B}\delta_{\nu x}+\hat{R}_{i\nu,B}\delta_{\mu x}-2\hat{R}_{i\mu,B}\hat{R}_{i\nu,B}\hat{R}_{ix,B}\right]\right\}\Delta r_{ix}. \quad (S9)$$

As discussed in the main text, for an interlayer vibrational mode in 2D materials, each layer vibrates as an almost rigid body and thus it can be simplified as a single object, where the structural details of each layer can be omitted. Subsequently, here $i$ indicates the index of an entire layer instead of any atom within it, and $B$ indicates a bond connecting from layer $i$ to a neighboring layer $i'$ in equilibrium. Recalling in the main text that the change of the polarizability by the shear vibration is $\Delta\alpha = \sum_i \alpha'_{ix}\Delta r_{ix}$, and $\alpha'_{ix}$ and $\Delta\alpha$ are second-rank tensors. Thus we have $\Delta\alpha_{\mu\nu} = \sum_i \alpha'_{ix,\mu\nu}\Delta r_{ix}$. Comparing this equation with the above Eq. S9, we arrive at

$$\alpha'_{ix,\mu\nu} = -\sum_{B}\left\{\hat{R}_{ix,B}\left[\frac{\alpha'_{\|,B}+2\alpha'_{\perp,B}}{3}\delta_{\mu\nu}+\left(\alpha'_{\|,B}-\alpha'_{\perp,B}\right)\left(\hat{R}_{i\mu,B}\hat{R}_{i\nu,B}-\frac{1}{3}\delta_{\mu\nu}\right)\right]\right.$$

$$\left.+\frac{\alpha_{\|,B}-\alpha_{\perp,B}}{R_{i,B}}\left[\hat{R}_{i\mu,B}\delta_{\nu x}+\hat{R}_{i\nu,B}\delta_{\mu x}-2\hat{R}_{i\mu,B}\hat{R}_{i\nu,B}\hat{R}_{ix,B}\right]\right\}. \quad (S10)$$

Similarly for an interlayer breathing mode, only the $z$ component of $\Delta\vec{r}_i(k)$ can be non-zero, and thus we obtain

$$\Delta\alpha_{\mu\nu} = -\sum_{iB}\left\{\hat{R}_{iz,B}\left[\frac{\alpha'_{\|,B}+2\alpha'_{\perp,B}}{3}\delta_{\mu\nu}+\left(\alpha'_{\|,B}-\alpha'_{\perp,B}\right)\left(\hat{R}_{i\mu,B}\hat{R}_{i\nu,B}-\frac{1}{3}\delta_{\mu\nu}\right)\right]\right.$$

$$\left.+\frac{\alpha_{\|,B}-\alpha_{\perp,B}}{R_{i,B}}\left[\hat{R}_{i\mu,B}\delta_{\nu z}+\hat{R}_{i\nu,B}\delta_{\mu z}-2\hat{R}_{i\mu,B}\hat{R}_{i\nu,B}\hat{R}_{iz,B}\right]\right\}\Delta r_{iz}. \quad (S11)$$

Again recalling in the main text that the change of the polarizability by the breathing vibration is $\Delta\alpha_{\mu\nu} = \sum_i \alpha'_{iz,\mu\nu}\Delta r_{iz}$. Comparing this equation with the above Eq. S11, we arrive at

$$\alpha'_{iz,\mu\nu} = -\sum_{B}\left\{\hat{R}_{iz,B}\left[\frac{\alpha'_{\|,B}+2\alpha'_{\perp,B}}{3}\delta_{\mu\nu}+\left(\alpha'_{\|,B}-\alpha'_{\perp,B}\right)\left(\hat{R}_{i\mu,B}\hat{R}_{i\nu,B}-\frac{1}{3}\delta_{\mu\nu}\right)\right]\right.$$

$$\left.+\frac{\alpha_{\|,B}-\alpha_{\perp,B}}{R_{i,B}}\left[\hat{R}_{i\mu,B}\delta_{\nu z}+\hat{R}_{i\nu,B}\delta_{\mu z}-2\hat{R}_{i\mu,B}\hat{R}_{i\nu,B}\hat{R}_{iz,B}\right]\right\}. \quad (S12)$$

Eq. S10 and Eq. S12 suggest that $\alpha'_{ix}$ or $\alpha'_{iz}$, the derivative of the system's polarizability with respect to the layer $i$'s displacement along the $x$ or $z$ direction, can be determined by the interlayer bond (length and direction), and bond polarizabilities.

According to Eq. S1, for the commonly used parallel polarization set-up in the backscattering geometry $z(xx)\bar{z}$, only the $xx$ components of the tensors need to be considered (i.e., $\mu = \nu = x$). Consequently, we have

$$\alpha'_{ix,xx} = -\sum_B \left\{ \frac{\alpha'_{\parallel,B} + 2\alpha'_{\perp,B}}{3}\hat{R}_{ix,B} + (\alpha'_{\parallel,B} - \alpha'_{\perp,B})\hat{R}^3_{ix,B} - \frac{\alpha'_{\parallel,B} - \alpha'_{\perp,B}}{3}\hat{R}_{ix,B} \right.$$

$$\left. + 2\frac{\alpha_{\parallel,B} - \alpha_{\perp,B}}{R_{i,B}}\hat{R}_{ix,B} - 2\frac{\alpha_{\parallel,B} - \alpha_{\perp,B}}{R_{i,B}}\hat{R}^3_{ix,B} \right\}$$

$$= -\sum_B \left\{ \frac{\alpha'_{\parallel,B} + 2\alpha'_{\perp,B}}{3} + (\alpha'_{\parallel,B} - \alpha'_{\perp,B})\hat{R}^2_{ix,B} - \frac{\alpha'_{\parallel,B} - \alpha'_{\perp,B}}{3} + 2\frac{\alpha_{\parallel,B} - \alpha_{\perp,B}}{R_{i,B}} - 2\frac{\alpha_{\parallel,B} - \alpha_{\perp,B}}{R_{i,B}}\hat{R}^2_{ix,B} \right\} \hat{R}_{ix,B}$$

$$= \sum_B C_{i,B}\hat{R}_{ix,B}, \tag{S13}$$

and

$$\alpha'_{iz,xx} = -\sum_B \left\{ \frac{\alpha'_{\parallel,B} + 2\alpha'_{\perp,B}}{3}\hat{R}_{iz,B} + (\alpha'_{\parallel,B} - \alpha'_{\perp,B})\hat{R}^2_{ix,B}\hat{R}_{iz,B} - \frac{\alpha'_{\parallel,B} - \alpha'_{\perp,B}}{3}\hat{R}_{iz,B} - 2\frac{\alpha_{\parallel,B} - \alpha_{\perp,B}}{R_{i,B}}\hat{R}^2_{ix,B}\hat{R}_{iz,B} \right\}$$

$$= -\sum_B \left\{ \frac{\alpha'_{\parallel,B} + 2\alpha'_{\perp,B}}{3} + (\alpha'_{\parallel,B} - \alpha'_{\perp,B})\hat{R}^2_{ix,B} - \frac{\alpha'_{\parallel,B} - \alpha'_{\perp,B}}{3} - 2\frac{\alpha_{\parallel,B} - \alpha_{\perp,B}}{R_{i,B}}\hat{R}^2_{ix,B} \right\} \hat{R}_{iz,B}$$

$$= \sum_B C^*_{i,B}\hat{R}_{iz,B}, \tag{S14}$$

where the coefficients $C_{i,B}$ and $C^*_{i,B}$ are related to the properties of the interlayer bond $B$ connecting from layer $i$ to a neighboring layer $i'$, such as the interlayer bond length and its $x$ component, and the interlayer bond polarizabilities and their radial derivatives.

It follows that the change of the polarizability is

$$\Delta\alpha_{xx} = \sum_i \alpha'_{ix,xx}\Delta r_{ix} \tag{S15}$$

by the shear vibrations and

$$\Delta\alpha_{xx} = \sum_i \alpha'_{iz,xx}\Delta r_{iz} \tag{S16}$$

by the breathing vibrations.

## S2. THE INTERLAYER BONDS AND POLARIZABILITY DERIVATIVES OF EACH LAYER

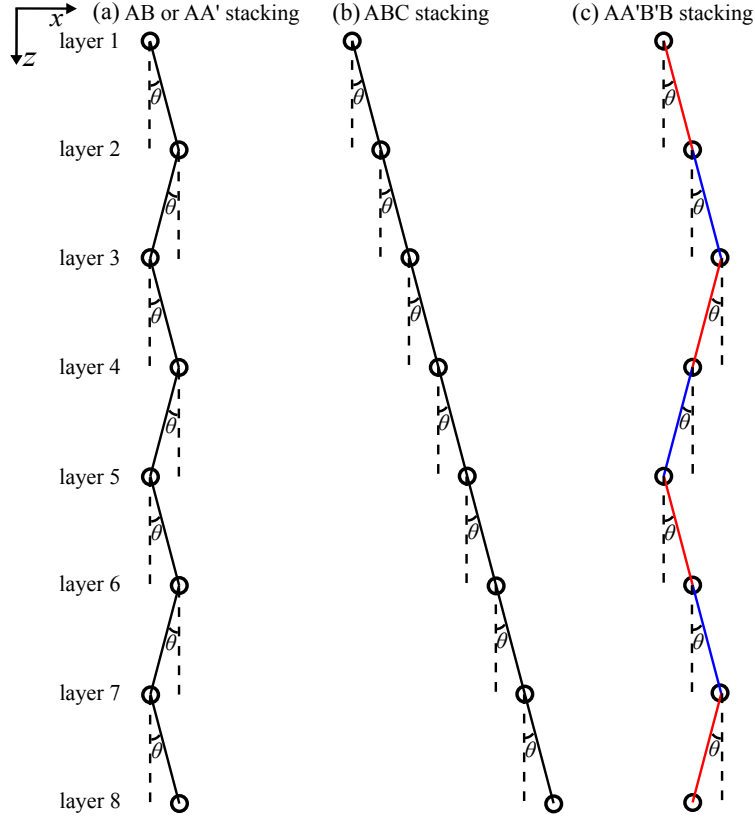

FIG. S2. Schematic of the interlayer bonds in $N$-layer for (a) AB or AA$'$ stacking, (b) ABC stacking, and (c) AA$'$B$'$B stacking types. Each layer is simplified as a single object for interlayer vibrations. AB or AA$'$ stacking repeats every two layer, ABC stacking repeats every three layers, and AA$'$B$'$B stacking repeats every four layers. In (c) AA$'$B$'$B stacking, the interlayer bonds corresponding to AA$'$ and AB stackings are differentiated by red and blue colors, respectively, indicating that AA$'$ and AB stackings alternate.

As shown in Figure S2, except that layer 1 and layer $N$ have only one interlayer bond, other interior layers $i$ have two interlayer bonds: one with the layer above $i-1$ and the other one with the layer below $i+1$. For an interior layer $i$, the $x$ components of these two normalized bond vectors assume a relation $\hat{R}_{ix,i-1} = \hat{R}_{ix,i+1}$ in (a) AB or AA$'$ stacking, while $\hat{R}_{ix,i-1} = -\hat{R}_{ix,i+1}$ in (b) ABC stacking. In addition, regardless of the stacking type, for layer $i$ and its neighboring layer $j$ ($i-1$ or $i+1$), there is a general relation $\hat{R}_{ix,j} = -\hat{R}_{jx,i}$.

For AB or AA$'$ stacking in Figure S2a, taking 6L as an example, $\hat{R}_{1x,2} = \sin\theta, \hat{R}_{2x,1} = \hat{R}_{2x,3} = -\sin\theta, \hat{R}_{3x,2} = \hat{R}_{3x,4} = \sin\theta, \hat{R}_{4x,3} = \hat{R}_{4x,5} = -\sin\theta, \hat{R}_{5x,4} = \hat{R}_{5x,6} = \sin\theta, \hat{R}_{6x,5} = -\sin\theta$. Thus the polarizability derivative with respect to the layer $i$'s displacement along the $x$ direction is $\alpha'_{1x,xx} = C\hat{R}_{1x,2} = C\sin\theta = \beta, \alpha'_{2x,xx} = C(\hat{R}_{2x,1} + \hat{R}_{2x,3}) = -2C\sin\theta = -2\beta, \alpha'_{3x,xx} = C(\hat{R}_{3x,2} + \hat{R}_{3x,4}) = 2C\sin\theta = 2\beta, \alpha'_{4x,xx} = -2\beta, \alpha'_{5x,xx} = 2\beta, \alpha'_{6x,xx} = C\hat{R}_{6x,5} = -C\sin\theta = -\beta$. Note that $C = C(AB)$ or $C = C(AA')$, the coefficient related to the interlayer bond polarizability and its derivatives in AB or AA' stacking, respectively. Taking 7L as an example, the interlayer bond vectors are not changed for layer 1 to layer 5, but layer 6 has $\hat{R}_{6x,5} = \hat{R}_{6x,7} = -\sin\theta$, while layer 7 has $\hat{R}_{7x,6} = \sin\theta$. Thus we have $\alpha'_{1x,xx} = \beta, \alpha'_{2x,xx} = -2\beta, \alpha'_{3x,xx} = 2\beta, \alpha'_{4x,xx} = -2\beta, \alpha'_{5x,xx} = 2\beta, \alpha'_{6x,xx} = C(\hat{R}_{6x,5} + \hat{R}_{6x,7}) = -2C\sin\theta = -2\beta, \alpha'_{7x,xx} = C\hat{R}_{7x,6} = C\sin\theta = \beta$. In general, for AB or AA' stacking, due to $\hat{R}_{ix,i-1} = \hat{R}_{ix,i+1}$, $\alpha'_{1x,xx} = \beta$, $\alpha'_{Nx,xx} = \beta$ for odd $N$ or $\alpha'_{Nx,xx} = -\beta$ for even $N$, and $\alpha'_{2x,xx} = -2\beta, \alpha'_{3x,xx} = 2\beta, \alpha'_{4x,xx} = -2\beta, \alpha'_{5x,xx} = 2\beta, ...$, where there is a repeated pattern of $-2\beta, 2\beta$ for the interior layers. Here $\beta = \beta_1$ for AA' stacking, while $\beta = \beta_2$ for AB stacking.

For ABC stacking in Figure S2b, due to $\hat{R}_{ix,i-1} = -\hat{R}_{ix,i+1}$, for an interior layer $i$, $\alpha'_{ix,xx} = C(\hat{R}_{ix,i-1} + \hat{R}_{ix,i+1}) = 0$, while for layer 1 and layer $N$, $\alpha'_{1x,xx} = C\hat{R}_{1x,2} = C\sin\theta = \beta$ and $\alpha'_{Nx,xx} = C\hat{R}_{Nx,N-1} = -C\sin\theta = -\beta$. Here $\beta = \beta_2$ for ABC stacking.

For AA'B'B stacking in Figure S2c, the periodicity corresponds to every four layers, and AA' and AB stackings (red and blue colors) alternate. The $x$ components of normalized interlayer bond vectors are $\hat{R}_{1x,2} = \sin\theta, \hat{R}_{2x,1} = -\hat{R}_{2x,3} = -\sin\theta, \hat{R}_{3x,2} = \hat{R}_{3x,4} = -\sin\theta, \hat{R}_{4x,3} = -\hat{R}_{4x,5} = \sin\theta, \hat{R}_{5x,4} = \hat{R}_{5x,6} = \sin\theta, \hat{R}_{6x,5} = -\hat{R}_{6x,7} = -\sin\theta, \hat{R}_{7x,6} = \hat{R}_{7x,8} = -\sin\theta, \hat{R}_{8x,7} = -\hat{R}_{8x,9} = \sin\theta, ...$, where an interior layer $i$ has the same interlayer bond vectors to layer $i+4$. Thus

$$\alpha'_{1x,xx} = C_{1,2}\hat{R}_{1x,2} = C(AA')\sin\theta = \beta_1,$$

$$\alpha'_{2x,xx} = C_{2,1}\hat{R}_{2x,1} + C_{2,3}\hat{R}_{2x,3} = -C(AA')\sin\theta + C(AB)\sin\theta = -\beta_1 + \beta_2,$$

$$\alpha'_{3x,xx} = C_{3,2}\hat{R}_{3x,2} + C_{3,4}\hat{R}_{3x,4} = -C(AB)\sin\theta - C(AA')\sin\theta = -\beta_2 - \beta_1,$$

$$\alpha'_{4x,xx} = C_{4,3}\hat{R}_{4x,3} + C_{4,5}\hat{R}_{4x,5} = C(AA')\sin\theta - C(AB)\sin\theta = \beta_1 - \beta_2,$$

$$\alpha'_{5x,xx} = C_{5,4}\hat{R}_{5x,4} + C_{5,6}\hat{R}_{5x,6} = C(AB)\sin\theta + C(AA')\sin\theta = \beta_2 + \beta_1,$$

$$\alpha'_{6x,xx} = C_{6,5}\hat{R}_{6x,5} + C_{6,7}\hat{R}_{6x,7} = -C(AA')\sin\theta + C(AB)\sin\theta = -\beta_1 + \beta_2,$$

$$\alpha'_{7x,xx} = C_{7,6}\hat{R}_{7x,6} + C_{7,8}\hat{R}_{7x,8} = -C(AB)\sin\theta - C(AA')\sin\theta = -\beta_2 - \beta_1,$$

$$\alpha'_{8x,xx} = C_{8,7}\hat{R}_{8x,7} + C_{8,9}\hat{R}_{8x,9} = C(AA')\sin\theta - C(AB)\sin\theta = \beta_1 - \beta_2,$$

$$\vdots$$

$$\alpha'_{Nx,xx} = \beta_1 (\text{if } N = 4m) \text{ or } \beta_2 (\text{if } N = 4m+1) \text{ or } -\beta_1 (\text{if } N = 4m+2) \text{ or } -\beta_2 (\text{if } N = 4m+3),$$

(S17)

where $m$ is an integer, $C(AA')\sin\theta = \beta_1$, and $C(AB)\sin\theta = \beta_2$. Note that for an interior layer $i$, $\alpha'_{ix,xx} = -\alpha'_{(i+2)x,xx}$, and thus $\alpha'_{ix,xx} = \alpha'_{(i+4)x,xx}$.

Turing to the $z$ direction (Figure S2), for an interior layer $i$, the $z$ components of the two normalized interlayer bond vectors always assume a relation $\hat{R}_{iz,i-1} = -\hat{R}_{iz,i+1}$ regardless of the in-plane stacking details. In addition, regardless of the stacking type, for layer $i$ and its neighboring layer $j$ ($i-1$ or $i+1$), there is a general relation $\hat{R}_{iz,j} = -\hat{R}_{jz,i}$. In AB or AA' or ABC stacking, for an interior layer $i$, the polarizability derivative with respect to its displacement along the $z$ direction is $\alpha'_{iz,xx} = C^*(\hat{R}_{iz,i-1} + \hat{R}_{iz,i+1}) = 0$, while for layer 1 and layer $N$, $\alpha'_{1z,xx} = C^*\hat{R}_{1z,2} = C^*\cos\theta = \gamma$ and $\alpha'_{Nz,xx} = C^*\hat{R}_{Nz,N-1} = -C^*\cos\theta = -\gamma$. Here $\gamma = \gamma_1$ for AA' stacking, while $\gamma = \gamma_2$ for AB or ABC stacking.

However, again for AA'B'B stacking, the situation is more complicated due to the mixture of AA' and AB stackings. In specific,

$$\alpha'_{1z,xx} = C^*_{1,2}\hat{R}_{1z,2} = C(AA')^*\cos\theta = \gamma_1,$$
$$\alpha'_{2z,xx} = C^*_{2,1}\hat{R}_{2z,1} + C^*_{2,3}\hat{R}_{2z,3} = -C(AA')^*\cos\theta + C(AB)^*\cos\theta = -\gamma_1 + \gamma_2,$$
$$\alpha'_{3z,xx} = C^*_{3,2}\hat{R}_{3z,2} + C^*_{3,4}\hat{R}_{3z,4} = -C(AB)^*\cos\theta + C(AA')^*\cos\theta = -\gamma_2 + \gamma_1,$$
$$\alpha'_{4z,xx} = C^*_{4,3}\hat{R}_{4z,3} + C^*_{4,5}\hat{R}_{4z,5} = -C(AA')^*\cos\theta + C(AB)^*\cos\theta = -\gamma_1 + \gamma_2,$$
$$\alpha'_{5z,xx} = C^*_{5,4}\hat{R}_{5z,4} + C^*_{5,6}\hat{R}_{5z,6} = -C(AB)^*\cos\theta + C(AA')^*\cos\theta = -\gamma_2 + \gamma_1,$$
$$\vdots$$
$$\alpha'_{Nz,xx} = -\gamma_1 \text{ (if } N = 2m) \text{ or } -\gamma_2 \text{ (if } N = 2m+1), \tag{S18}$$

where $m$ is an integer, $C(AA')^*\cos\theta = \gamma_1$, and $C(AB)^*\cos\theta = \gamma_2$. Note that for an interior layer $i$, $\alpha'_{iz,xx} = -\alpha'_{(i+1)z,xx}$ and thus $\alpha'_{iz,xx} = \alpha'_{(i+2)z,xx}$.

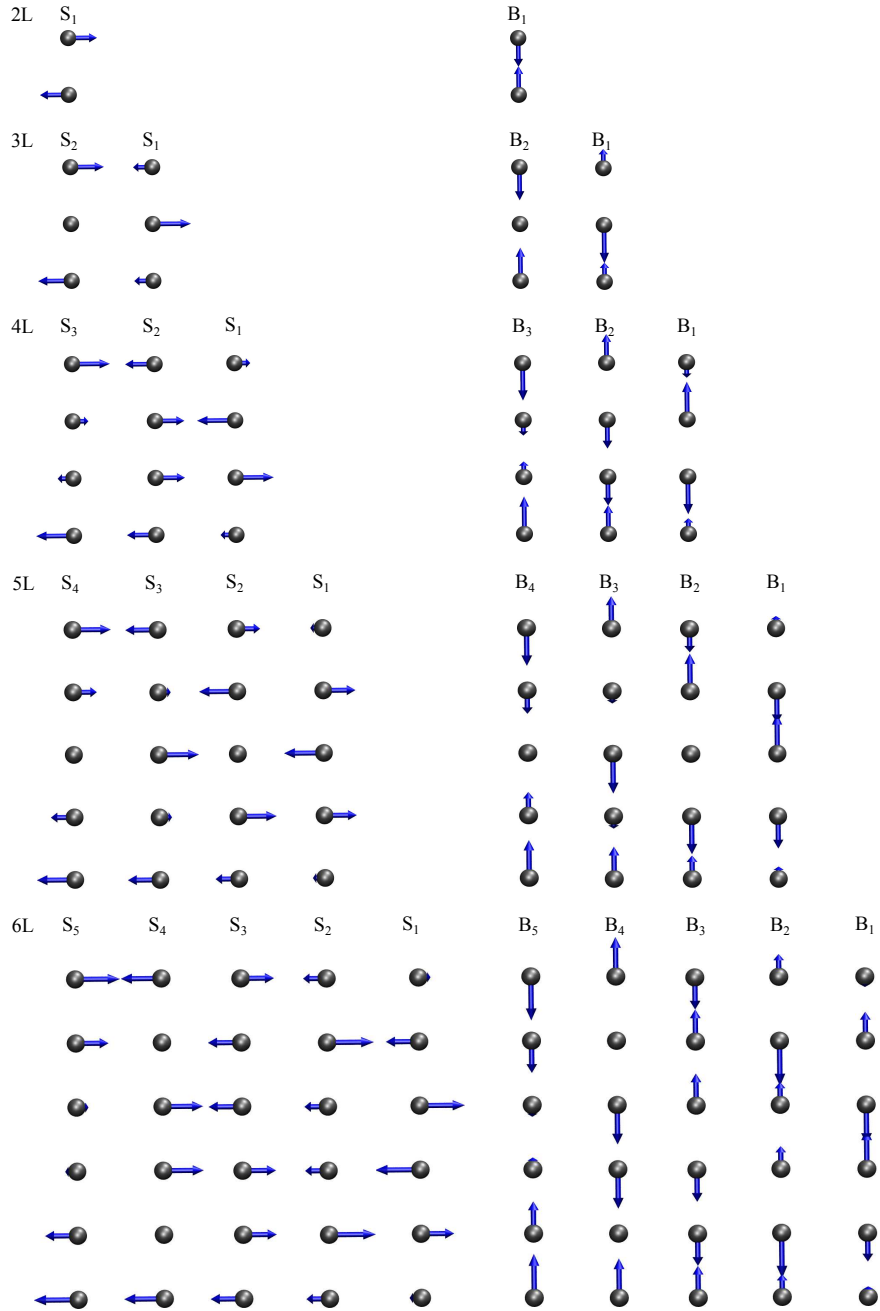

FIG. S3. Schematic of the vibrations of the interlayer shear (S) and breathing (B) modes from 2L to 6L, derived from the linear chain model. For these interlayer vibrations, each layer is treated as a single object (the gray sphere), and the blue arrows indicate both the direction and magnitude of the vibrations of each layer. For $N$L, there are $N-1$ S and B modes, where $S_1$ ($B_1$) is the highest-frequency S (B) mode, while $S_{N-1}$ ($B_{N-1}$) is the lowest-frequency S (B) mode.

## S3. PARAMETER FITTING FOR TRILAYER MoSe$_2$

As discussed in the main text, for trilayer MoSe$_2$ in AA$'$B$'$ stacking, the polarizability change by the shear vibrations can be derived as follows:

$$\Delta\alpha_{xx}(\text{AA}'\text{B}', \text{S}_2) = \frac{\beta_1 + \beta_2}{\sqrt{2}}; \qquad \Delta\alpha_{xx}(\text{AA}'\text{B}', \text{S}_1) = \sqrt{1.5}(\beta_1 - \beta_2).$$

For 3L MoSe$_2$, the frequencies of the S$_2$ and S$_1$ modes are $\omega_2 \approx 14$ cm$^{-1}$ and $\omega_1 \approx 23$ cm$^{-1}$, respectively. In AA$'$B$'$ stacking, we have $I(\text{S}_2) = \frac{n_2+1}{\omega_2}|\Delta\alpha_{xx}(\text{AA}'\text{B}', \text{S}_2)|^2 = 0.5\frac{n_2+1}{\omega_2}|\beta_1 + \beta_2|^2 \approx 0.59|\beta_1 + \beta_2|^2$ and $I(\text{S}_1) = \frac{n_1+1}{\omega_1}|\Delta\alpha_{xx}(\text{AA}'\text{B}', \text{S}_1)|^2 = 1.5\frac{n_1+1}{\omega_1}|\beta_1 - \beta_2|^2 \approx 0.61|\beta_1 - \beta_2|^2$. Here $n_i = (e^{\hbar\omega_i/k_BT} - 1)^{-1}$ is the Bose-Einstein distribution of phonon occupation at room temperature $T = 300$K. Based on the experimental Raman data of bilayer MoSe$_2$,[6] we know $|\beta_1|/|\beta_2| = 2.32$ as mentioned in the main text.

If $\beta_1$ and $\beta_2$ are assumed to be real variables as in the common non-resonant Raman modeling, then $\beta_1 = 2.32\beta_2$ or $\beta_1 = -2.32\beta_2$. For the former case, we have $I(\text{S}_2) = 0.59|\beta_1 + \beta_2|^2 = 0.59|2.32\beta_2 + \beta_2|^2 = 6.50|\beta_2|^2$, and $I(\text{S}_1) = 0.61|\beta_1 - \beta_2|^2 = 0.61|2.32\beta_2 - \beta_2|^2 = 1.06|\beta_2|^2$, thereby giving $I(\text{S}_2)/I(\text{S}_1) = 6.13$; for the latter case, we have $I(\text{S}_2) = 0.59|\beta_1 + \beta_2|^2 = 0.59|-2.32\beta_2 + \beta_2|^2 = 1.03|\beta_2|^2$, and $I(\text{S}_1) = 0.61|\beta_1 - \beta_2|^2 = 0.61|-2.32\beta_2 - \beta_2|^2 = 6.72|\beta_2|^2$, thereby giving $I(\text{S}_2)/I(\text{S}_1) = 0.15$. Both cases yield very unequal intensities of the S2 and S1 modes, which are consistent with first-principles non-resonant Raman calculations in Ref. [6].

However, the S$_2$ and S$_1$ modes exhibited nearly equal intensities in the experimental resonant Raman scattering.[6] In reality, the polarizability (or dielectric function) has both real and imaginary parts due to the light absorption in experimental resonant Raman scattering.[7,8] Thus $\beta_1$ and $\beta_2$ are complex variables: $\beta_1 = |\beta_1|e^{i\phi_1}; \beta_2 = |\beta_2|e^{i\phi_2}$, where $\phi_1$ and $\phi_2$ are their phase angles, respectively. To have $I(\text{S}_2) = I(\text{S}_1)$, we need $0.59|\beta_1 + \beta_2|^2 = 0.61|\beta_1 - \beta_2|^2$, which is

$$|\beta_1 + \beta_2|^2 = 1.034|\beta_1 - \beta_2|^2 \longrightarrow$$
$$|\beta_1|^2 + |\beta_2|^2 + \beta_1\beta_2^* + \beta_1^*\beta_2 = 1.034\left(|\beta_1|^2 + |\beta_2|^2 - \beta_1\beta_2^* - \beta_1^*\beta_2\right) \longrightarrow$$
$$2.034(\beta_1\beta_2^* + \beta_1^*\beta_2) = 0.034\left(|\beta_1|^2 + |\beta_2|^2\right) \longrightarrow$$
$$2.034|\beta_1||\beta_2|\left(e^{i(\phi_1-\phi_2)} + e^{-i(\phi_1-\phi_2)}\right) = 0.034\left(|\beta_1|^2 + |\beta_2|^2\right) \longrightarrow$$
$$4.068|\beta_1||\beta_2|\cos(\phi_1 - \phi_2) = 0.034\left(|\beta_1|^2 + |\beta_2|^2\right) \longrightarrow$$
$$\cos(\phi_1 - \phi_2) = 0.008\frac{|\beta_1|^2 + |\beta_2|^2}{|\beta_1||\beta_2|} \tag{S19}$$

With $|\beta_1| = 2.32|\beta_2|$, we arrive at $\cos(\phi_1 - \phi_2) = 0.022$, which yields $|\phi_1 - \phi_2| \approx 88.74°$. This suggests that for AA′ and AB stackings, their complex interlayer bond polarizabilities and derivatives not only have different magnitudes, but also have different phase angles in the resonant Raman scattering. Here we assume $|\beta_1| = 2.32$ and $\phi_1 = 118.74°$, while $|\beta_2| = 1.00$ and $\phi_2 = 30.00°$ without loss of generality. These parameters give rise to nearly equal intensities between the $S_2$ and $S_1$ modes for trilayer MoSe$_2$ in AA′B′ stacking.